\pdfoutput=1
\documentclass[12pt]{iopjournal}

\usepackage{amsmath, amssymb, amsfonts}
\usepackage{booktabs}
\usepackage[numbers,sort&compress]{natbib}

\hypersetup{
    linkcolor=blue!70!black,
    filecolor=magenta,
    urlcolor=cyan!70!black,
    citecolor=red!70!black
}

\def\ben{\begin{equation}}
\def\een{\end{equation}}
\newcommand{\mr}[1]{\mathrm{#1}}
\newcommand{\mc}[1]{\mathcal{#1}}
\newcommand{\mb}[1]{\mathbf{#1}}
\newcommand{\ud}{\mr{d}}
\newcommand{\ui}{\mr{i}}

\renewcommand{\articletype}[1]{}
\renewcommand{\headrulewidth}{0pt}

\begin{document}

\articletype{Paper}

\title{Observables and anti-Hermitian generators in time-dependent unitary coupled cluster theory}

\author{Mart\'in A. Mosquera}

\affil{Department of Chemistry and Biochemistry, Montana State University, Bozeman, Montana
59717, USA}

\email{martinmosquera@montana.edu}


\begin{abstract}
This work presents a time-dependent (TD) unitary coupled-cluster (UCC) formulation for electronic
quantum dynamics, including the propagation of excited states and their superpositions, in both
single- and multi-reference regimes. Standard TD coupled-cluster techniques offer size-extensivity,
but they rely on non-Hermitian bivariational action functionals that break time-reversibility,
giving transition matrix elements and amplitude estimators that are asymmetric (though accurate
and systematically improvable). Here we use the
time-evolution operator as an exponential map driven by TD anti-Hermitian cluster operators and
first-order generators. Applying the Dirac-Frenkel action principle, we extract equations of motion
governed by Heisenberg-picture-like commutators. This approach connects to our previous
non-Hermitian formulations, where observables are expressed in terms of regular and extended cluster
operators. From that connection we obtain a generator cluster operator whose unperturbed TD limit
leads to the UCC eigenvalue problem. Even though this problem can be exact, the time dependence of
the generator holds only at short propagation times, so we use the generator to prepare the initial
state and then propagate that state with the formal TD UCC equation of motion.  The theory is tested
on an extended hard-core Bose-Hubbard ring with connections to neutral atom chains, and we discuss
the theory's present limitations and possible extensions.
\end{abstract}

\section{Introduction}

The systematically improvable computation of real-time quantum dynamics is essential to understand
the behavior of strongly correlated systems driven out of equilibrium. Examples of these phenomena
span attosecond photochemistry, non-adiabatic multi-state dynamics, and the coherent control of
molecular materials \cite{krausz2014attosecond, lepine2014attosecond, nisoli2017attosecond,
palacios2020quantum, nelson2020non, matsika2021electronic, casida2012progress, mosquera2015time,
kirrander2020ehrenfest, fernandez2016non, li2005ab, mosquera2016sequential}. Quantum
Information Science (QIS) has matured around controllable hardware platforms such as
superconducting circuits, photonic networks, and trapped ions
\cite{kjaergaard2020superconducting, flamini2018photonic, bruzewicz2019trapped, wang2020integrated}.
Programmable neutral atom clusters and Rydberg atom arrays are also emerging as
platforms for quantum computation and analog many-body quantum simulation
\cite{saffman2016quantum, bernien2017probing, browaeys2020many, bluvstein2022quantum,
henriet2020quantum, weiss2017quantum, wurtz2023aquila, brennen1999quantum, jaksch1999entanglement,
jaksch2000fast, ahn2000information, omran2019generation, gross2017quantum, evered2023high,
saffman2019quantum, levine2018high, graham2019rydberg, saffman2010quantum, saffman2008scaling,
ates2007many, graham2022multi}. These neutral atom systems offer fine spatial control and long
coherence times, but putting their multi-reference configurations to practical use
relies on a microscopic understanding of their coherent, real-time evolution under
external electromagnetic driving forces.

For accurate stationary ground-state simulations, Coupled-Cluster (CC) theory is a standard in
\textit{ab initio} quantum chemistry due to its size-extensivity capabilities and elegant treatment
of many-body correlation \cite{cizek1966correlation, bartlett2007coupled, Paldus1994,
vcivzek1969use, shavitt1998history, bartlett1978many, helgaker2014molecular}. For excited states,
frameworks such as the symmetry-adapted-cluster CI \cite{nakatsuji1978cluster,
nakatsuji1979cluster}, Linear Response CC, and Equation-of-Motion CC obtain spectra by diagonalizing
the non-Hermitian similarity-transformed Hamiltonian \cite{monkhorst1977calculation,
koch1990coupled, christiansen1998response, stanton1993equation, krylov2008equation,
bartlett2012coupled, sneskov2012excited, koch1991analytical, pedersen1997coupled,
nascimento2019general}. Separately, extensive developments in Multireference CC (MRCC) have enabled
accurate targeting of strongly correlated ground and excited states
\cite{mukherjee1989multireference, evangelista2018perspective, jeziorski2010multireference,
mahapatra1998state, mahapatra1999size, evangelista2011orbital, hanauer2012communication,
hanrath2005exponential, hanrath2008multi, datta2012multireference, park2020multireference,
maitra2012unitary, kohn2013state, chattopadhyay2000development, samanta2014excited, jagau2012linear,
lechner2021perturbative,hirata2026}. The real-time dynamics of these systems under external time-dependent
perturbations is a challenging problem. Standard Time-Dependent CC (TD-CC) methods rely on the
bivariational principle introduced by Arponen, Dalgaard, and Monkhorst \cite{arponen1983variational,
dalgaard1983some}. Because the similarity transformation is non-unitary, the bra and ket states must
be parameterized as distinct, independent degrees of freedom. This has nonetheless led to powerful
approaches and applications \cite{kvaal2012ab, pedersen2019symplectic, sato2018communication,
skeidsvoll2020time, skeidsvoll2022simulating, white2018time, white2019time, pathak2021time,
pathak2022time, koulias2019relativistic, wang2022accelerating, sverdrup2023time, pedersen1998time,
pedersen2020interpretation, pigg2012time, hoodbhoy1978time, hoodbhoy1979time, schonhammer1978time,
guha1991multireference, nascimento2016linear, nascimento2017simulation}.  Standard TD-CC yields
transition matrix elements that break time-reversibility, sacrificing formal Hermitian symmetry
\cite{harsha2018difference}. In some cases, restoring this property comes at the expense of losing
some other desirable aspect \cite{kats2011second, walz2012application}. 

To avoid these asymmetries and restore physical Hermiticity, one can rely on unitary coupled-cluster
(UCC) theory \cite{kutzelnigg1977pair, prasad1985some, bartlett1989unitary, kutzelnigg1991error,
taube2006new}. Standard ground-state UCC restricts the cluster operators to be anti-Hermitian; as a
consequence, the sum-over-states polarization propagator and transition moments feature the correct
symmetric properties \cite{taube2006new}. UCC has a long history, including self-consistent electron
propagator theory \cite{prasad1985some, liu2018unitary} and property calculations without quantum
phase estimation \cite{hodecker2020unitary}, but its resurgence is tied to its being a source of
\textit{ansatz} for chemistry algorithms formulated for
quantum computing \cite{cao2019quantum, mcardle2020quantum, bauer2020quantum, motta2022emerging,
ryabinkin2018qubit, romero2018strategies, xia2020qubit, tilly2020computation, anand2022quantum,
shen2017quantum, lee2018generalized, evangelista2019exact, higgott2019variational,
sokolov2020quantum, cervia2021lipkin, smith2019approaching}. For hardware, current circuit depths
demand approximations such as nonunitary extensions via midcircuit measurements
\cite{fleury2024nonunitary}, or quantum variants of the Davidson algorithm for excited states
\cite{kim2023two}. The outstanding difficulty in UCC is the non-terminating
Baker-Campbell-Hausdorff (BCH) expansion and the implied non-commutative algebra. Classical
implementations have worked around it with automated non-antisymmetric tensor contractions
\cite{datta2013non}. The internally contracted EOM (equation-of-motion) UCC formalism effectively 
truncates the series off at the four-fold commutator level \cite{li2025equation}; there the excited states are built from a
forward-only excitation operator and obtained as a stationary eigenvalue problem, rather than as the
time-dependent superpositions propagated here. Downfolding methods such as the Double Unitary
Coupled-Cluster formalism take a different route, partitioning space and orbitals to decouple
internal degrees of freedom \cite{bauman2019downfolding, kowalski2020time}.

Alongside these developments, the author has pursued a sequence of extensions to standard TD-CC
aimed at propagating quantum superpositions efficiently while preserving size-extensivity. These
are excited-state response theory within the CC framework \cite{mosquera2022excited,
mosquera2021second}, Second Response Theory \cite{mosquera2023second}, and extended Time-Dependent
Multireference Coupled-Cluster theory \cite{mosquera2025}. They treat superpositions as
infinitesimal bivariational perturbations, resembling the generators of continuous group
theory. The present work investigates the non-commutative approach to the static and time-dependent
properties of quantum systems, time-reversibility, and symmetric transition matrix elements in both
regimes. The main TD equation for reference propagation follows from the Dirac-Frenkel action and
the Maurer-Cartan relations, from which we extract the equation of motion for the generator of
excitations. Analyzing its unperturbed dynamics gives the EOM-UCC eigenvalue problem; factoring out
unlinked ``vacuum'' fluctuations leaves a symmetric eigenvalue equation.  The solution then provides
initial operators that ``rotate'' the reference, leading to a general equation of motion. Exact
numerical simulations of a neutral atom ring model show that the formulation stays well-behaved
under mid-to-strong correlation conditions.

\section{Theory}

\subsection{Correlated Ground State}
This work concerns propagation under non-relativistic Hamiltonians of the kind
$\hat{H}(t)=\hat{H}_0+\hat{V}(t)$, where $\hat{H}_0$ is the static component and $\hat{V}(t)$
represents an external driving TD potential. The operator $\hat{H}_0$, as free Hamiltonian, is used
to obtain the eigen-spectrum of the system.

We denote the reference state as $|0\rangle$, which can be constructed through the solution of a
multi- or single-reference model. In previous work \cite{mosquera2025}, we used this state to
construct the basis excitation operators, $\{\tau_{\bar{\mu}}\}$, through solution of a SVD problem,
followed by a QR factorization so this set of operators is similar to the single-reference
formalism. This construction slightly differs from standard orthogonalization in internally
contracted MRCC~\cite{hanauer2012communication}, which places the external excitations on a
contracted footing, and removes the purely internal excitations, which are replaced by orbital
relaxation. The SVD discards null directions from the overlap metric (singular
values below a threshold $\eta$), while the subsequent QR factorization rotates the reference
projection onto a single axis, fixing $\hat{\tau}_0 = \hat{1}$ with $\hat{\tau}_0|0\rangle =
|0\rangle$, and yielding $\{|\bar{\mu}\rangle = \hat{\tau}_{\bar{\mu}}|0\rangle\}_{\bar{\mu}\ge 0}$
as an orthonormal set. The multireference system depends on a single orthonormal excitation ladder,
which makes the formalism below resemble a single-reference one. However, the basis excitation
operators do not commute. For theoretical purposes we assume that
$[\hat{\tau}_{\bar{\mu}},\hat{\tau}_{\bar{\nu}}]\neq 0$ in general, unless the system is single-reference or
bosonic as in the numerical application, section \ref{sec:implementation} (in
these cases non-commutativity remains present, but only between excitation operators and their
conjugate forms). Throughout this work, the $\hat{\tau}_{\bar{\mu}}$ are excitation operators in the
general sense. The formalism below uses only their algebra, their mutual non-commutativity, and the
action of their adjoints on the reference, not the statistics of the underlying particles. Therefore
the formalism applies to fermionic and bosonic systems. We use it here to study a system of
hard-core bosons, Sec.~\ref{sec:bose_hubbard}.

The ground state wavefunction is expressed using the anti-Hermitian operator $\hat{T}$:
\begin{equation}
    |\Psi_0\rangle \equiv \exp(\hat{T})|0\rangle~, \quad \text{with} \quad \hat{T}^\dagger = -\hat{T}~.
\end{equation}
Because $\hat{T}$ is anti-Hermitian, the ground state $|\Psi_0\rangle$ is normalized ($\langle
\Psi_0 | \Psi_0 \rangle = 1$). Expectation values evaluated over the reference are
denoted as $\langle \hat{O} \rangle = \langle 0 | \hat{O} | 0 \rangle$. For example, we average
ground-state observables by
evaluating the transformed operators over the reference determinant, $\langle \Psi_0 | \hat{O} |
\Psi_0 \rangle = \langle e^{-\hat{T}} \hat{O} e^{\hat{T}} \rangle \equiv \langle \hat{O}_T \rangle$.
As expected, the ground state problem is solved by minimizing the energy function $\mathcal{E}(\hat{T})=\langle
\hat{H}_T\rangle$, leading to the stationary condition for the ground state
\begin{equation}
    \langle [\hat{H}_T,\hat{\pi}_T(\hat{\kappa}_{\mu})]\rangle = 0
\end{equation}
for all anti-Hermitian excitation basis operators $\hat{\kappa}_{\mu}$ (defined next, section \ref{sec:gs_propagation}), 
which lets us write $\hat{T}=\sum_{\mu}t_{\mu}\hat{\kappa}_{\mu}$. The
superoperator $\hat{\pi}_T$ is expressed as
\begin{equation}
    \hat{\pi}_T \cdot = 1 - \frac{1}{2}[\hat{T}, \cdot] + \frac{1}{3!}[\hat{T}, [\hat{T}, \cdot]] - \dots = \sum_{n=0}^\infty \frac{(-1)^n}{(n+1)!} \mathrm{ad}_{\hat{T}}^n \cdot~,
\end{equation}
where the adjoint operation is $\mathrm{ad}_{\hat{T}} \cdot = [\hat{T}, \cdot]$.

\subsection{Ground State Propagation}\label{sec:gs_propagation}
We write the time-evolution operator as $\hat{U}_{\mr{m}}(t)$, acting on $|0\rangle$ using an
anti-Hermitian cluster operator:
\begin{equation}
    \hat{U}_{\mathrm{m}}(t; g)|0\rangle = \exp\left[ \hat{x}_{\mathrm{m}}(t; g) \right]|0\rangle~.
\end{equation}
The propagation operator $\hat{x}_{\mr{m}}$ depends on both time and an auxiliary parameter $g$,
which is related to continuous generators (section \ref{sec:extended_generators}). 
To track real-time transitions, we define an
anti-Hermitian basis. We introduce an extended index $\mu \equiv (\bar{\mu}, \sigma)$ that combines
the conventional excitation index $\bar{\mu}$ and a type index $\sigma \in \{-, +\}$. The
corresponding anti-Hermitian basis operators are defined as:
\begin{equation}\label{eq:basis_def}
    \hat{\kappa}_{\mu} = \begin{cases} 
    \hat{\tau}_{\bar{\mu}} - \hat{\tau}_{\bar{\mu}}^\dagger & \text{for } \sigma = - \quad \text{(Real antisymmetric)} \\
    \ui(\hat{\tau}_{\bar{\mu}} + \hat{\tau}_{\bar{\mu}}^\dagger) & \text{for } \sigma = + \quad \text{(Imaginary symmetric)}
    \end{cases}
\end{equation}
Conventional CC restricts basis expansions to excitation operators ($\hat\tau_{\bar\mu}$). Each
$\hat\kappa_\mu$ above is anti-Hermitian by construction. The real-antisymmetric and
imaginary-symmetric sectors let the generator span the full excitation space while preserving the
anti-Hermiticity.

The cluster operator $\hat{x}_{\mathrm{m}}(t; g)$ is constructed as a single linear expansion over
this extended non-scalar index:
\begin{equation}
    \hat{x}_{\mathrm{m}}(t; g)|0\rangle = \sum_{\mu} x_{\mathrm{m},\mu}(t; g)
    \hat{\kappa}_{\mu}|0\rangle~.
\end{equation}
$x_{\mathrm{m},\mu}(t;g)$ is the amplitude associated to the $\hat{\kappa}_{\mu}$ basis state.  The
basis operators in the expanded set are anti-Hermitian ($\hat{\kappa}_{\mu}^\dagger =
-\hat{\kappa}_{\mu}$), the generator itself remains anti-Hermitian at all times
($\hat{x}_{\mathrm{m}}^\dagger = -\hat{x}_{\mathrm{m}}$), ensuring that $\hat{U}_{\mr{m}}^\dagger(t)
\hat{U}_{\mr{m}}(t) = \mb{1}$.

To find the equations of motion, we evaluate the Dirac-Frenkel action over the reference state
$|0\rangle$:
\begin{equation}\label{eq:action}
    \mc{S} = \int \ud t~\langle \hat{U}_{\mr{m}}^\dagger(t) [\ui\partial_t - \hat{H}(t)]
    \hat{U}_{\mr{m}}(t) \rangle~.
\end{equation}
The dynamics are determined by the condition $\delta\mc{S}=0$. Since $\hat{U}_{\mr{m}}(t)$ is
unitary, the wave-function is normalized in a robust way. Using differentiation rules for
exponential matrices on the state $|\Psi(t)\rangle = \hat{U}_{\mr{m}}(t) |0\rangle$, we obtain
\begin{equation}
    \hat{U}_{\mr{m}}^\dagger \ui \partial_t \hat{U}_{\mr{m}} = \ui \hat{\pi}_{x_{\mathrm{m}}}(\partial_t \hat{x}_{\mathrm{m}}(t))~,
\end{equation}
where the projection superoperator evaluates to the infinite nested operator commutator expansion:
\begin{equation}\label{eq:gamma_rate}
    \hat{\pi}_{x_{\mathrm{m}}}(\partial_t \hat{x}_{\mathrm{m}}(t)) = \sum_{n=0}^\infty
    \frac{(-1)^n}{(n+1)!} \mathrm{ad}_{\hat{x}_{\mathrm{m}}(t)}^n \big( \partial_t
    \hat{x}_{\mathrm{m}}(t) \big)~, 
\end{equation}
we have that $\mathrm{ad}_{\hat{x}_{\mathrm{m}}(t)}\hat{O}=[\hat{x}_{\mr{m}}(t),\hat{O}]$.
We used this type of expansion above in previous non-Hermitian work \cite{mosquera2025}.
\sloppy Now, let us define the fully transformed Hamiltonian $\hat{H}_{x_{\mathrm{m}}}(t) =
\exp[-\hat{x}_{\mathrm{m}}(t)]\hat{H}(t)\exp[\hat{x}_{\mathrm{m}}(t)]$, and the residual operator:
\begin{equation}\label{eq:total_residual}
    \hat{R}_{\mathrm{m}}(t; g) = \ui\hat{\pi}_{x_{\mathrm{m}}}(\partial_t \hat{x}_{\mathrm{m}}(t;
    g)) - \hat{H}_{x_{\mathrm{m}}}(t; g)~.
\end{equation}
Variations of the action over $\hat{x}_{\mathrm{m}}$ are taken along the transformed basis
directions $\hat{\Xi}_{\mathrm{m},\mu}(t; g) \equiv \hat{\pi}_{x_{\mathrm{m}}}(\hat{\kappa}_\mu)$,
i.e., the projection superoperator of Eq.~(\ref{eq:gamma_rate}) applied to the basis operators
$\hat{\kappa}_\mu$ of Eq.~(\ref{eq:basis_def}). This leads to the
equations of motion:
\begin{equation}\label{eq:base_eom}
    \langle \left[ \hat{\Xi}_{\mathrm{m},\mu}(t; g) \, , \, \hat{R}_{\mr{m}}(t; g) \right] \rangle = 0~.
\end{equation}
For $g=0$ the above gives the equation that propagates the ground state.

\subsection{Generators of Excitations}\label{sec:extended_generators}

As in our previous formulations, we examine a generator of excitations/de-excitations. 
We expand $\hat{x}_{\mr{m}}(t)$ with respect to a continuous parameter $g$:
\begin{equation}\label{eq:generator_expansion}
    \hat{x}_{\mathrm{m}}(t; g) = \hat{x}(t) + g \hat{z}(t) + \mathcal{O}(g^2)~.
\end{equation}
The operator $\hat{x}(t)$ describes the non-linear propagation of the ground state
$|\Psi_0\rangle$ under the external field (or $\hat{V}(t)$), while $\hat{z}(t)$ denotes the first-order
generator. The initial condition for $\hat{x}(t)$ is:
\begin{equation}
    \hat{x}(0) = \hat{T}~.
\end{equation}
A weak driving field can lead the system to build such
a contribution from an anti-Hermitian operator, giving the generator ($\hat{z}$) a physical origin.

For $g=0$ the residual reads $\hat{R}_x(t) = \ui\hat{\pi}_x (\partial_t \hat{x}(t)) - \hat{H}_x(t)$, 
where $\hat{H}_x(t)$ is $\exp[-\hat{x}(t)]\hat{H}(t)\exp[\hat{x}(t)]$.
Before we extract the EOM for the generator, we differentiate the residual $\hat{R}_{\mathrm{m}}$
with respect to $g$, and define the continuous generator $\tilde{z}(t; g) =
\hat{\pi}_{x_{\mathrm{m}}}(\partial_g \hat{x}_{\mathrm{m}}(t; g))$, the projected form of the
first-order generator $\hat{z}(t)$ of Eq.~(\ref{eq:generator_expansion}). Evaluating this at $g=0$
leads to the additional definition $\tilde{z}(t) = \tilde{z}(t; g)\big|_{g=0}$.

Now we invoke the Maurer-Cartan (MC) relation (detailed in SI):
\begin{equation}\label{eq:trick}
    \partial_g \left( \hat{\pi}_{x_{\mathrm{m}}}(\partial_t \hat{x}_{\mathrm{m}}) \right) = \partial_t \tilde{z}(t; g) + \left[ \hat{\pi}_{x_{\mathrm{m}}}(\partial_t \hat{x}_{\mathrm{m}}) \, , \, \tilde{z}(t; g) \right]~.
\end{equation}
The $g$-derivative of the residual thus reads:
\begin{equation}
    \partial_g \hat{R}_{\mathrm{m}}(t; g) = \ui\partial_t \tilde{z}(t; g) + [\hat{R}_{\mathrm{m}}(t; g), \tilde{z}(t; g)]~.
\end{equation}
Evaluating at $g=0$, where $\hat{R}_{\mathrm{m}}\big|_{g=0} =
\hat{R}_x(t)$, yields the term:
\begin{equation}
    \hat{R}^{(1)}(t) = \ui\partial_t \tilde{z}(t) + [\hat{R}_x(t), \tilde{z}(t)]~. \label{eq:R1}
\end{equation}
The basis variations $\hat{\Xi}_{\mu}^{(n)}(t) = \partial_g^n \hat{\Xi}_{\mathrm{m},\mu}(t; g)
\big|_{g=0}$ are extracted via their respective Lie derivatives (e.g., $\partial_g
\hat{\Xi}_{\mathrm{m},\mu} = \partial_\mu \tilde{z}(t; g) + [\hat{\Xi}_{\mathrm{m},\mu},
\tilde{z}(t; g)]$, with $\partial_\mu \equiv \partial/\partial x_{\mathrm{m},\mu}$), giving:
\begin{align}
    \hat{\Xi}_\mu^{(0)}(t) &= \hat{\pi}_x(\hat{\kappa}_\mu)~, \label{eq:xi_0} \\
    \hat{\Xi}_\mu^{(1)}(t) &= \partial_\mu \tilde{z}(t) + [\hat{\Xi}_\mu^{(0)}(t), \tilde{z}(t)]~. \label{eq:xi_1}
\end{align}
These are the elements needed to write the dynamical equations for $\tilde{z}(t)$.

The EOM for the ground-state trajectory ($g=0$) requires the term $\hat{\Xi}_\mu^{(0)}(t)$, and the
relation $\langle \left[
\hat{\Xi}_{\mu}^{(0)}(t) \, , \, \hat{R}_x(t) \right] \rangle = 0$.
Differentiating Eq.~(\ref{eq:base_eom}) with respect to $g$ and evaluating at $g=0$ gives the constraint:
\begin{equation}
    \langle \left[ \hat{\Xi}_{\mu}^{(1)}(t) \, , \, \hat{R}_x(t) \right] \rangle + \langle \left[ \hat{\Xi}_{\mu}^{(0)}(t) \, , \, \hat{R}^{(1)}(t) \right] \rangle = 0~.
\end{equation}
Substituting the evaluation for the first-order residual $\hat{R}^{(1)}(t)$ into this constraint
provides the initial-value equation, where the
first-order TD equation for the generator becomes:
\begin{equation}\label{eq:z1_eom}
    \langle \left[ \hat{\Xi}_{\mu}^{(0)}(t) \, , \, \ui\partial_t \tilde{z}(t) \right] \rangle = - \langle \left[ \hat{\Xi}_{\mu}^{(0)}(t) \, , \, [\hat{R}_x(t), \tilde{z}(t)] \right] \rangle - \langle \left[ \hat{\Xi}_\mu^{(1)}(t) \, , \, \hat{R}_x(t) \right] \rangle~.
\end{equation}
As shown later on, this relation applies to free and driven propagations, and is accurate at
early times.

\subsection{Unperturbed Non-Stationary States, Initial Conditions, and Eigenvalue Problem}\label{sec:unperturbed_states}

Propagating a general initial state requires cluster operators at $t=0$. We obtain them
from the unperturbed non-stationary states, which evolve without
introducing spurious frequencies \cite{mosquera2023second}. The unperturbed equation for
$\tilde{z}(t)$ yields the EOM-UCC eigenvectors and energies, with $\hat{x}(t) = \hat{T}$ and
$\partial_t \hat{x} = 0$. The residual operator then becomes $\hat{R}_x(t) = -\hat{H}_T$, where $\hat{H}_T
= e^{-\hat{T}}\hat{H}_0 e^{\hat{T}}$. Inserting it into Eq.~(\ref{eq:z1_eom})
gives:
\begin{equation}\label{eq:z1_unperturbed_eom_full}
    \langle \left[ \hat{\Xi}_{0,\mu} \, , \, \ui\partial_t \tilde{z}(t) - [\hat{H}_T, \tilde{z}(t)] \right] \rangle = \langle \left[ \hat{\Xi}_\mu^{(1)}(t) \, , \, \hat{H}_T \right] \rangle~,
\end{equation}
where $\hat{\Xi}_{0,\mu} = \hat{\pi}_T(\hat{\kappa}_\mu)$ is a transformed static basis object.
The right-hand side of Eq.~(\ref{eq:z1_unperturbed_eom_full}) is zero, as long as one can express
$\hat{\Xi}_{\mu}^{(1)}(t)$ in the basis $\{\hat{\Xi}_{0,\mu}\}$ (which demands algebraic
closedness), because ground-state
optimization imposes $\langle [\hat{\Xi}_{0,\mu}, \hat{H}_T] \rangle = 0$.

Hence, the unperturbed first-order EOM simplifies to:
\begin{equation}\label{eq:z1_homogeneous_eom}
    \langle \left[ \hat{\Xi}_{0,\mu} \, , \, \ui\partial_t \tilde{z}(t) - [\hat{H}_T, \tilde{z}(t)] \right] \rangle = 0~.
\end{equation}
This equation requires an oscillatory solution. Since the generator $\tilde{z}(t)$ is
anti-Hermitian ($\tilde{z}^\dagger(t) = -\tilde{z}(t)$), the fluctuation operator
cannot be a single complex exponential; the general solution combines forward
($e^{-\ui\Omega_I t}$) and backward ($e^{\ui\Omega_I t}$) frequencies. We take the eigen-vectors to
be
\begin{equation}
    \tilde{X}^I = \sum_\mu X_\mu^I \hat{\Xi}_{0,\mu}~.
\end{equation}
The coefficients $X_{\mu}^I$ are complex-valued, so this operator is not anti-Hermitian.
Expanding over this basis is convenient as it brings in a metric matrix consistent with the way the
solutions to the Dirac-Frenkel problem are treated.

In standard non-Hermitian TD-MRCC, transition elements involve numerical shifts because there are
left/right wavefunctions that are different for the same quantum state.  The unitary framework
can eliminate these. With $\hat{T}$ anti-Hermitian, the Hamiltonian $\hat{H}_T = e^{-\hat{T}}\hat{H}_0
e^{\hat{T}}$ remains Hermitian. Averaging the EOM-UCC relation
$[\hat{H}_T, \tilde{X}^I] = \Omega_I \tilde{X}^I$ over the reference gives:
$
    0 = \langle [\hat{H}_T, \tilde{X}^I] \rangle = \Omega_I \langle \tilde{X}^I \rangle~.
$
For regular transitions with $\Omega_I \neq 0$, this gives $\langle
\tilde{X}^I \rangle = 0$, which is exact in the closed-algebra cases, and FCI (full
configuration interaction) limits. Otherwise $\langle
\tilde{X}^I \rangle$ could be small, but nonzero. Although $\langle \hat{\kappa}_\mu \rangle =
0$ by construction of the $\{\hat{\tau}_{\bar{\mu}}\}$ operators, the operators
$\hat{\Xi}_{0,\mu} = \hat{\pi}_T(\hat{\kappa}_\mu)$ involve nested commutators with $\hat{T}$,
that is, products of these operators whose reference expectations reduce to overlaps. 
Because the eigenvalue problem and the equations of motion imply
only commutators, each $\tilde{X}^I$ is defined up to an additive multiple of the
identity. One can fix this freedom, for example, through the convention
\begin{equation}\label{eq:mode_convention}
    \tilde{X}^I \leftarrow \tilde{X}^I - \langle \tilde{X}^I \rangle \hat{1}~,
\end{equation}
with the adjoint vectors carrying the conjugate constant; both constants drop from every commutator
below and are kept implicit hereafter. The constant induced in the generator $\tilde{z}(0)$
(defined below) by this shift is purely imaginary, so anti-Hermiticity is preserved. 
With this convention,
$\langle \tilde{X}^I \rangle = 0$ holds at any truncation order. We invoke this convention 
below to make statements about $\langle \tilde{z}(0)
\rangle$ exact; the calculations reported here do not apply it
(Sec.~\ref{sec:bose_hubbard}), but it is important for supermolecular additivity, as discussed in
the supplementary material. In any case, the condition $\langle \tilde{X}^I\rangle=0$ is stronger
than Eq.~(\ref{eq:mode_convention}). Now,
since the operators $\hat{\Xi}_{0,\mu}$ are anti-Hermitian
($\hat{\Xi}_{0,\mu}^{\dagger} = -\hat{\Xi}_{0,\mu}$), the adjoint of a given eigenvector is:
\begin{equation}
    \tilde{X}^{I\dagger} = \sum_\mu (X_\mu^I)^* \hat{\Xi}_{0,\mu}^{\dagger} = - \sum_\mu (X_\mu^I)^* \hat{\Xi}_{0,\mu}~.
\end{equation}
These relations yield the unperturbed expression:
\begin{equation}\label{eq:z1_unperturbed_solution_homo}
    \tilde{z}(t) = \sum_{I>0} \left( c_I e^{-\ui\Omega_I t} \tilde{X}^I - c_I^* e^{\ui\Omega_I t} \tilde{X}^{I\dagger} \right)~.
\end{equation}
The operator $\tilde{z}(t)$ is anti-Hermitian at general truncation orders.

Substituting this expansion into Eq.~(\ref{eq:z1_homogeneous_eom}) and grouping the terms with
$e^{-\ui\Omega_I t}$, we find the EOM-UCC eigenvalue problem:
\begin{equation}\label{eq:practical_eigenvalue}
    \sum_\nu \langle \left[ \hat{\Xi}_{0,\mu} \, , \, \left[ \hat{H}_T, \hat{\Xi}_{0,\nu} \right] \right] \rangle X_\nu^I = \Omega_I \sum_\nu \langle \left[ \hat{\Xi}_{0,\mu} \, , \, \hat{\Xi}_{0,\nu} \right] \rangle X_\nu^I~.
\end{equation}
The terms with $e^{\ui\Omega_I t}$ lead to the conjugate problem:
\begin{equation}\label{eq:practical_eigenvalue_conjugate}
    -\Omega_I \langle \left[ \hat{\Xi}_{0,\mu} \, , \, \tilde{X}^{I\dagger} \right] \rangle = \langle \left[ \hat{\Xi}_{0,\mu} \, , \, \left[ \hat{H}_T, \tilde{X}^{I\dagger} \right] \right] \rangle~.
\end{equation}
The practical matrix eigenvalue problem reads $\mathbf{A}\mathbf{X}^I = \Omega_I
\mathbf{S}\mathbf{X}^I$. The Jacobian matrix is $A_{\mu\nu} = \langle [ \hat{\Xi}_{0,\mu},
[\hat{H}_T, \hat{\Xi}_{0,\nu}] ] \rangle$, and the metric matrix is $S_{\mu\nu} = \langle [
\hat{\Xi}_{0,\mu}, \hat{\Xi}_{0,\nu} ] \rangle$. 
Using the Jacobi identity over the commutators gives:
\begin{equation}
    A_{\mu\nu} - A_{\nu\mu} = \langle [ \hat{\Xi}_{0,\mu}, [\hat{H}_T, \hat{\Xi}_{0,\nu}] ] \rangle - \langle [ \hat{\Xi}_{0,\nu}, [\hat{H}_T, \hat{\Xi}_{0,\mu}] ] \rangle = +\langle [ \hat{H}_T, [\hat{\Xi}_{0,\mu}, \hat{\Xi}_{0,\nu}] ] \rangle~.
\end{equation}
If the commutators close, the right-hand side vanishes, as happens in an FCI-style
calculation or under $\mathfrak{su}(M)$ adaptation; otherwise one symmetrizes the Jacobian by hand,
$\mb{A}\leftarrow (1/2)(\mb{A}+\mb{A}^{\dagger})$.

With either route, the Jacobian is symmetric ($A_{\mu\nu} = A_{\nu\mu}$), and the
metric matrix anti-symmetric ($S_{\mu\nu} = -S_{\nu\mu}$). The projection can then be carried out
with the adjoint vector $\tilde{X}^{I\dagger}$, so no separate left-hand operators are needed.
Evaluated at $t=0$, the unperturbed solution gives the initial condition for $\tilde{z}$:
\begin{equation}\label{eq:z1_initial_condition}
    \tilde{z}(0) = \sum_{I>0} \left( c_I \tilde{X}^I - c_I^* \tilde{X}^{I\dagger} \right)~.
\end{equation}
The coefficients $\{c_I\}$ are the linear (complex) amplitudes of the initial state.

\subsection{Propagation of Linear Superpositions}

Let us consider $\hat{U}_{\mathrm{QM}}(t)$ as the conventional full quantum mechanical evolution
operator, $\hat{U}_{\mathrm{QM}}(t)=\mathcal{T}\exp(-\mr{i}\int_0^t\hat{H}(s)\mathrm{d}s)$.
Its action is reproduced by a CC propagator, denoted $\hat{U}_y(t)$, which unlike the full QM
operator depends on the initial state.
The generator of the previous subsection acts like a rotation of the reference.
Consider a general coherent initial state, expressed in terms of a (complex) ground-state amplitude $c_0$
and a static, anti-Hermitian operator $\tilde{z}(0)$ acting upon the correlated UCC ground state.
The resulting propagated state is written as: 
\begin{equation}\label{eq:psi_t}
\begin{split}
    |\Psi(t)\rangle &= \hat{U}_{\mathrm{QM}}(t)e^{\hat{T}}[c_0 +
    \tilde{z}(0)]|0\rangle~,\\
    & = \hat{U}_{y}(t)[c_0 + \tilde{z}(0)]|0\rangle~,
\end{split}
\end{equation}
$\tilde{z}(0)$ describes the excited-state superposition component of the overall initial state;
$c_0=0$ is a valid value.
Because $\hat{U}_{y}(t)$ describes the time dependence, observables follow from the Heisenberg-like
operator $\tilde{B}_{y}(t) =
\hat{U}_{y}^\dagger(t) \hat{B} \hat{U}_{y}(t)$, giving:
\begin{equation}\label{eq:heisenberg_b}
    \langle \hat{B} \rangle(t) = \frac{ \langle [c_0^* - \tilde{z}(0)] \tilde{B}_{y}(t) [c_0 +
    \tilde{z}(0)] \rangle }{ |c_0|^2 + \langle -\tilde{z}^2(0) \rangle }~.
\end{equation}
The minus sign is only a reflection of Hermiticity, but norms are strictly positive. The denominator is
the squared norm $\langle \bar{\Psi}(0) | \bar{\Psi}(0) \rangle$
($|\bar{\Psi}(0)\rangle=[c_0+\tilde{z}(0)]|0\rangle$), where the cross terms
$c_0^* \langle \tilde{z}(0) \rangle - c_0 \langle \tilde{z}(0) \rangle$ vanish identically.
$\tilde{z}(0)$ is a sum of nonzero-frequency EOM-UCC vectors $\tilde{X}^I$ and their adjoints, and
the reference-component convention of Eq.~(\ref{eq:mode_convention}) enforces $\langle \tilde{X}^I
\rangle = 0$, hence $\langle \tilde{z}(0) \rangle = 0$ (Sec.~\ref{sec:unperturbed_states}). 
Because $\tilde{z}(0)$ is static, these observables should not
diverge, which can be seen if the above is rewritten as
\begin{align}\label{eq:size_extensive_heisenberg}
    \langle \hat{B} \rangle(t) &= \langle \tilde{B}_{y}(t) \rangle + \frac{1}{|c_0|^2 +
    \langle -\tilde{z}^2(0) \rangle} \Big( c_0^* \langle \tilde{B}_{y}(t) \tilde{z}(0)
    \rangle - c_0 \langle \tilde{z}(0) \tilde{B}_{y}(t) \rangle \nonumber \\
    &\quad  + \frac{1}{2} \langle [[\tilde{B}_{y}(t), \tilde{z}(0)], \tilde{z}(0)]\rangle
- \frac{1}{2} \langle \{ \tilde{B}_{y}(t), \tilde{z}^2(0) \}
    \rangle_{\mr{C}} \Big)~.
\end{align}
Only benign commutators and their ``anti'' forms appear. The normalization
factor is formally present because we normalize with respect to the metric matrix of the EOM-UCC
problem.

The propagator is given by a time-dependent, anti-Hermitian cluster operator
$\hat{y}(t)$, with $\hat{U}_{y}(t) = \exp[\hat{y}(t)]$,
$\hat{y}(t=0)=\hat{T}$, and no parameter $g$ employed in this case. Applying the Dirac-Frenkel
action principle to this superposition reference yields the working equation 
(it reduces to Eq.~(\ref{eq:base_eom}) when $\tilde{z}(0)=0$):
\begin{equation}\label{eq:superposition_eom}
    \langle \bar{\Psi}(0) | \left[ \hat{\Xi}_{y,\mu}(t) \, , \, \hat{R}_y(t) \right] | \bar{\Psi}(0) \rangle = 0~.
\end{equation}
The transformed basis element reads $\hat{\Xi}_{y,\mu}(t) = \hat{\pi}_{y(t)}(\hat{\kappa}_\mu)$, and
the residual is $\hat{R}_y(t) = \ui\hat{\pi}_{y}(\partial_t \hat{y}(t)) - \hat{H}_{y}(t)$, with
$\hat{H}_{y}(t) = e^{-\hat{y}(t)}\hat{H}(t)e^{\hat{y}(t)}$. 

Finally, let us assume that $\hat{H}(-t) = \hat{H}(t)$. The generator is
anti-Hermitian at all times,
\begin{equation}\label{eq:anti_hermitian_y}
    \hat{y}^\dagger(t) = -\hat{y}(t)~,
\end{equation}
so $\hat{U}_y(t) = e^{\hat{y}(t)}$ is unitary. In Eq.~(\ref{eq:superposition_eom}), 
the sign flips of $\ui$ and of $\partial_{-t}$ cancel in
$\ui\hat{\pi}_y(\partial_t\hat{y})$, so complex conjugation combined with $t \to -t$
leads to a solution generated by the conjugated reference,
\begin{equation}\label{eq:time_reversal_y}
    \big[ \hat{y}_{\bar{\Psi}}(t) \big]^{*} = \hat{y}_{\bar{\Psi}^{*}}(-t)~,
\end{equation}
with the subindex indicating conjugation of the starting reference state.
The initial condition is the same because $\hat{T}^{*} = \hat{T}$. 
For a real observable $\hat{B}$, the numerator of
Eq.~(\ref{eq:heisenberg_b}) is real by Eq.~(\ref{eq:anti_hermitian_y}) and is carried
to the conjugated reference by Eq.~(\ref{eq:time_reversal_y}), while the denominator
is a norm, so
\begin{equation}\label{eq:observable_even}
    \langle \hat{B} \rangle_{\bar{\Psi}}(-t) = \langle \hat{B} \rangle_{\bar{\Psi}^{*}}(t)~.
\end{equation}
As long as the references are related by a global phase, $|\bar{\Psi}^{*}(0)\rangle
\propto |\bar{\Psi}(0)\rangle$, the two sides relate to each other and
Eqs.~(\ref{eq:time_reversal_y}) and (\ref{eq:observable_even}) reduce to
$\hat{y}(-t) = \hat{y}^{*}(t)$ and $\langle \hat{B} \rangle(-t) =
\langle \hat{B} \rangle(t)$.

\section{Numerical Simulation Details}\label{sec:implementation}

\subsection{Bose-Hubbard Model and Interaction Picture Solution}\label{sec:bose_hubbard}

\begin{figure}[htp!]
\centering
\includegraphics[scale=1.5]{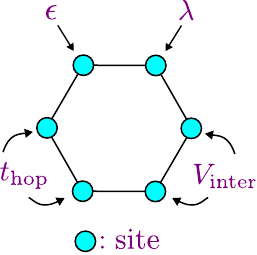}
\caption{Sketch of the model, a six-site ring ($L=6$) with inter-site coupling ($V_{\mr{inter}}$), on-site
energy ($\epsilon$), on-site linear source field strength ($\lambda$), and inter-site hopping
($t_{\mr{hop}}$). Each site carries a hard-core boson and is therefore in a ``0'' or ``1'' state,
mimicking a neutral-atom cluster.}
\label{fig:model}
\end{figure}

To implement the equations of motion computationally, we consider a finite-dimensional matrix
representation \cite{mosquera2022excited, mosquera2025}, and define a 1D hard-core bosonic
model, depicted in Fig.~\ref{fig:model}. The time-dependent Hamiltonian reads:
\begin{equation}
    \hat{H}(t) = \hat{H}_0 - f(t)\hat{\mu}~,
\end{equation}
where $\hat{H}_0$ spans $L$ interacting sites arranged on a ring:
\begin{equation}
    \hat{H}_0 = \epsilon \sum_{j=1}^L \hat{n}_j
    + V_{\mr{inter}} \sum_{j=1}^{L} \hat{n}_j \hat{n}_{j+1}
    + t_{\mathrm{hop}} \sum_{j=1}^{L} \left( \hat{b}_j^\dagger \hat{b}_{j+1} + \hat{b}_{j+1}^\dagger \hat{b}_j \right)
    + \lambda \sum_{j=1}^L \left( \hat{b}_j^\dagger + \hat{b}_j \right)~.
\end{equation}
The operators $\hat{b}_j^\dagger$ and $\hat{b}_j$ are the creation and annihilation operators at
site $j$, and $\hat{n}_j = \hat{b}_j^\dagger \hat{b}_j$ (the boson number operator); each site can
hold at most one boson. These operators follow $(\hat{b}_j)^2=(\hat{b}_j^{\dagger})^2=0$,
$[\hat{b}_i,\hat{b}_j^{\dagger}]=\delta_{ij}(1-2\hat{n}_j)$, and
$[\hat{b}_i,\hat{b}_j]=[\hat{b}_i^{\dagger},\hat{b}_j^{\dagger}]=0$ for all $i,j$.
This numerical application thus exercises non-commutativity between $\hat{\tau}_{\bar{\mu}}^{\dagger}$ and
$\hat{\tau}_{\bar{\nu}}$ (implying non-commutativity among $\hat{\kappa}_{\mu}$ objects or
$\hat{\Xi}_{0,\mu}$ objects), but not between pairs like $(\hat{\tau}_{\bar{\mu}},~\hat{\tau}_{\bar{\nu}})$ or
$(\hat{\tau}_{\bar{\mu}}^{\dagger},~\hat{\tau}_{\bar{\nu}}^{\dagger})$, $\bar{\mu}\neq\bar{\nu}$. These properties prevent
natural truncation of the BCH series. 

The on-site energy is $\epsilon$, $t_{\mr{hop}}$ is the hopping parameter, $V_{\mr{inter}}$ the
nearest-neighbor repulsion term, and $\lambda$ is an on-site source field strength. Periodic
boundary conditions are imposed as $\hat{b}_{L+1} \equiv \hat{b}_1$, so that the interaction and
hopping terms close the chain into a ring. The term involving $\lambda$ breaks boson-number
conservation (or excitation conservation).  The constant shift of the EOM vectors,
Eq.~(\ref{eq:mode_convention}), serves a theoretical purpose, but it is not required for the
calculations below. This point is further discussed below and in the supplementary material. There
we also show details about the reference construction, which is constructed by including the
no-boson state and all the single-boson configurations. The reference state $|0\rangle$ is built by
diagonalizing the static Hamiltonian $\hat{H}_0$ over those configurations and selecting $|0\rangle$
as the lowest-energy state.

It is convenient to switch to the interaction picture. To this end we write $|\Psi(t)\rangle =
e^{-\mr{i}\hat{H}_0t}\hat{U}_{\mr{I}}(t)[c_0+\tilde{z}(0)]|0\rangle$, and define
$\hat{V}_{\mr{I}}(t) = e^{\mr{i}\hat{H}_0t}\hat{V}(t)e^{-\mr{i}\hat{H}_0t}$, where $\hat{V}(t) =
-f(t)\hat{\mu}$. The UCC operator reads $\hat{U}_{\mr{I}}(t)=\exp[\hat{y}_{\mr{i}}(t)]$, with
$\hat{y}_{\mr{i}}(t=0) = \hat{T}$. The new ``$\Xi$'' term is
$\hat{\Xi}_{\mr{I},\mu}(t)=\hat{\pi}_{y_{\mr{i}}}(\hat{\kappa}_{\mu})$. Based on these definitions,
and the expansion $\hat{y}_{\mr{i}}(t)=\sum_{\mu}y_{\mr{i},\mu}(t)\hat{\kappa}_{\mu}$, the numerical
equation to solve takes the form:
\begin{equation}
\sum_{\nu} \mr{Im}[S_{\mu\nu}'(t)]\dot{y}_{\mr{i},\nu} = -\mr{Re}[h_{\mr{eff},\mu}(t)]~,
\end{equation}
where $S_{\mu\nu}' = \langle
\bar{\Psi}(0)|\hat{\Xi}_{\mr{I},\mu}(t)\hat{\Xi}_{\mr{I},\nu}(t)|\bar{\Psi}(0)\rangle$, and
$h_{\mr{eff},\mu}(t) =\langle \bar{\Psi}(0)|
\hat{\Xi}_{\mr{I},\mu}(t)e^{-\hat{y}_{\mr{i}}(t)}\hat{V}_{\mr{I}}(t)e^{\hat{y}_{\mr{i}}(t)}|\bar{\Psi}(0)\rangle$.
Observables are evaluated using Eq.~(\ref{eq:heisenberg_b}) after setting $\hat{U}_y(t) =
e^{-\mr{i}\hat{H}_0t}e^{\hat{y}_{\mr{i}}(t)}$. Our formulation is based on Heisenberg-type matrix
mechanics, with all operators represented in the $\{|\bar{\mu}\rangle\}$ basis. The exponential maps
and $\hat{\pi}$ superoperators are evaluated using the spectral theorem and its functional calculus.
These avoid truncation errors, and are efficient for the system treated here.

We apply a Gaussian pulse to study the time-evolution of the system. For $t\ge 0$, the time-dependent field is:
\begin{equation}
    f(t)\hat{\mu} = f_0 \exp\left(-\frac{(t-t_0)^2}{2\sigma^2}\right) \mu_0 \sum_{j=1}^L \left( \hat{b}_j^\dagger + \hat{b}_j \right)~.
\end{equation}
For a numerical simulation over a ring of $L = 6$ sites, we use the following parameters: $\epsilon = 1.0$
eV, $V_{\mr{inter}} = 0.500$ eV, $t_{\mathrm{hop}} = 0.050$ eV, $\lambda = 0.200$ eV (equivalently
$0.03675$, $0.01838$, $0.001838$, and $0.00735$ a.u.), $\mu_0 = 0.25$ a.u., $f_0 =
0.25$ a.u., $\sigma = 50$ a.u., and $t_0 = 150$ a.u. The simulation covers the time interval $t \in
[0, 500]$ a.u.

\subsection{Piecewise Propagation}\label{sec:piecewise}

Under strong driving, continuous propagation of the cluster operator $\hat{y}_{\mathrm{i}}(t)$ can drive
the amplitudes to large values. In a truncated UCC space, heavily displaced generators amplify the
out-of-subspace truncation errors inherent to the BCH expansion. To keep the interaction-picture
amplitudes bounded, and prevent ``degradation'' of the metric as well, we introduce a piecewise
restart protocol that re-centers the reference frame. Consider the interaction-picture state vector
at an absolute propagation time $t$, Eq.~(\ref{eq:psi_t}). When the integration reaches a boundary
time $\tau$, the state is evaluated as $|\Psi(\tau)\rangle = e^{-\mathrm{i}\hat{H}_0 \tau}
\hat{U}_{y_{\mathrm{i}}}(\tau) |\bar{\Psi}(0)\rangle$. At this point, the history of the wave packet
is absorbed into an updated reference state,
$
    |\bar{\Psi}(0)\rangle \leftarrow e^{-\mathrm{i}\hat{H}_0 \tau} \hat{U}_{y_{\mathrm{i}}}(\tau) |\bar{\Psi}(0)\rangle~.
$
The time-independent reference Hamiltonian is then shifted to include the external potential at $t =
\tau$,
$
    \hat{H}_0 \leftarrow \hat{H}_0 + \hat{V}(\tau)~.
$
To strictly preserve the total physical Hamiltonian, $\hat{H}(t) = \hat{H}_0^{(\mathrm{new})} +
\hat{V}^{(\mathrm{new})}(t)$, the time-dependent interaction potential for the subsequent interval must
be redefined as the residual fluctuation:
$
    \hat{V}(t) \leftarrow \hat{V}(t) - \hat{V}(\tau)~,
$
so that the effective perturbation starts from zero in the new epoch ($\hat{V}(\tau) = 0$).

The local time coordinate is reset by $t \leftarrow t
- \tau$. Because the new reference $|\bar{\Psi}(0)\rangle$ is the physical state at the boundary,
the dynamic cluster operator for the new interval must be the identity, $
\hat{y}_{\mathrm{i}}(0^{+}) = 0~. $ For propagation in the new time slot, the state is evaluated
using the new relative time $t$ as $ |\Psi(t)\rangle = e^{-\mathrm{i}\hat{H}_0 t}
\hat{U}_{y_{\mathrm{i}}}(t) |\bar{\Psi}(0)\rangle~, $ where $\hat{H}_0$, $\hat{V}(t)$, and
$|\bar{\Psi}(0)\rangle$ carry their newly updated values. In an untruncated space or closed
algebraic system, the BCH expansion is closed, while the piecewise evaluation matches the continuous
single-exponential ansatz. Otherwise, the consecutive piecewise layers generate higher-order,
out-of-basis terms through the nested commutators of the BCH series.

\section{Results and Discussion}

\subsection{Numerical}

\begin{figure}[htp!]
\centering
\includegraphics[scale=0.55]{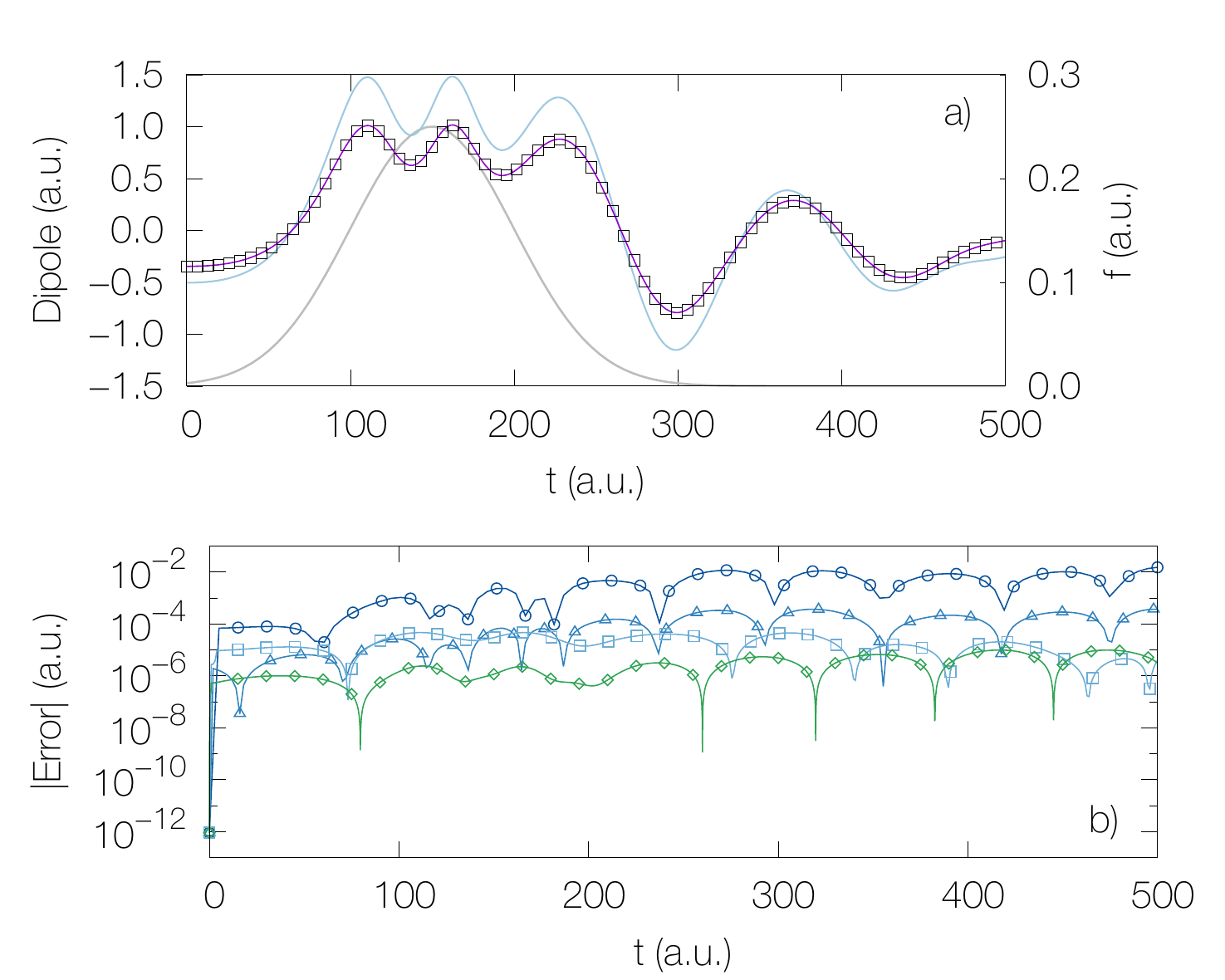}
\caption{a) Time-dependent dipole of the LC-initiated state (dark violet line, TDUCC; black open
squares, FCI), together with the ground-state-initiated dipole $\mu_{00}$ (light blue)
and, on the right-hand axis, the Gaussian driving field $f(t)$ (grey). b) Absolute deviation
$|\mu^{\mr{UCC}}(t) - \mu^{\mr{FCI}}(t)|$ of the LC-initiated dipole, on a logarithmic scale, for
four step counts over the same window $t \in [0, 500]$ a.u.: $N_t = 100$ (dark blue, open circles),
$250$ (medium blue, open triangles), $500$ (light blue, open squares) and $5000$ (green, open
diamonds).}
\label{fig:stacked_dipoles}
\end{figure}

We test the TD-UCC approach on the hard-core boson model of Sec.~\ref{sec:bose_hubbard}. For the
driven dipole, the state populations, a dipole-dipole correlation function, and the
reference-excited coherence, the deviation from the exact propagation for fine time-step simulations
stays at or below $10^{-5}$, against signals of order $10^{-1}$. The two-time correlation function
is a quantity examined here for which the theory is not exact; we treat it separately below.  The
following discussion focuses on expectation values (static matrix elements such as
$\langle\Psi_I|\hat{B}|\Psi_J\rangle$ can be deduced from Eq.~(\ref{eq:size_extensive_heisenberg}),
as discussed in the supplementary material).

Figure~\ref{fig:stacked_dipoles} shows the dipole under the external driving field. Panel (a) gives
the non-linear dipole dynamics obtained by propagating a linear combination (LC) superposition
state, with $c_0=1/\sqrt{17}$ and $\tilde{z}(0) = c_1(\tilde{X}^1-\tilde{X}^{1\dagger})$, $c_1 =
4/\sqrt{17}$; the ground-state-initiated dipole (from $e^{\hat{T}}|0\rangle$) is shown as a
reference curve and the Gaussian driving field $f(t)$ appears on the right-hand axis. The initial
superposition produces a non-stationary response of amplitude $0.35$ a.u., which the unitary
coupled-cluster framework reproduces in comparison to the FCI
propagation. Figure~\ref{fig:stacked_dipoles}(b) gives the absolute deviation between the UCC and
FCI dipoles for the LC-initiated state at four step counts over the same $500$ a.u.\ window.

We observe behavior common to RK4 integrators, with an exception. Going from $N_t = 100$ to $250$ steps divides the
peak error by $41$, close to the $39$ expected from $(2.5)^4$, lowering from $1.5 \times 10^{-2}$ to
$3.6 \times 10^{-4}$ a.u. Beyond $N_t \approx 10^3$ the scaling stops. The time-averaged error
settles around the order of $10^{-6}$ a.u., where a fivefold refinement for $N_t = 5000$ makes the
peak error improve by less than a factor of $1.5$. The error is not uniform within a single
propagation; it collapses at the nodes of the response, down to $1 \times 10^{-9}$ a.u.\ against a
peak of $1 \times 10^{-5}$ a.u. The instantaneous error value spans more than four orders of
magnitude. Such behavior is familiar in TD unitary theories. During the RK4 propagation the rates
$\dot{y}_{\mr{i},\mu}$ follow from the real linear system $\sum_{\nu} \mr{Im}[S_{\mu\nu}'(t)]\,
\dot{y}_{\mr{i},\nu} = -\mr{Re}[h_{\mr{eff},\mu}(t)]$. Depending on the driven state and the active
basis, the coefficient matrix $\mr{Im}[S(t)]$ can be subject to ill conditioning. Increasing the
number of steps is not sufficient to reduce further the error. After the propagation starts the
metric matrix can introduce numerical noise. In this case, we use complete orthogonal decomposition
for the solution of the linear problem. With solver pivot tolerance of $10^{-8}$ the RK4 accuracy
does not improve much beyond 500 timesteps, whereas with $10^{-9}$ we see a more systematic error
reduction. However, further reducing the COD threshold does not improve the error and can revert
the trend, evidencing the expected issue with the metric matrix. These results suggest that such
matrix needs to be well characterized for robust propagations based on it. There are potentially
multiple pathways to improve stability, such as rotation of the reference (but these are unexplored
in this work).

Regarding Fig.~\ref{fig:stacked_dipoles}(a), the ground-state-initiated and LC-initiated dipoles
follow similar oscillation profiles, despite the very different initial states. The reason is
connected to the couplings and translational symmetry. The dipole $\hat{\mu} = \mu_0 \sum_{j=1}^L
(\hat{b}_j^\dagger + \hat{b}_j)$ is a spatially uniform, single collective coordinate, and the
driving field couples to it alone, which is related to the generalized Kohn theorem. The ground
state and the LC superposition differ only in the internal interactions (due to $V_{\mr{inter}}$),
which $\hat{\mu}$ does not capture, so the two dipole signals reduce to a similar collective
response, separated by an approximate amplitude prefactor. 

\begin{figure}[htp!]
\centering
\includegraphics[scale=0.55]{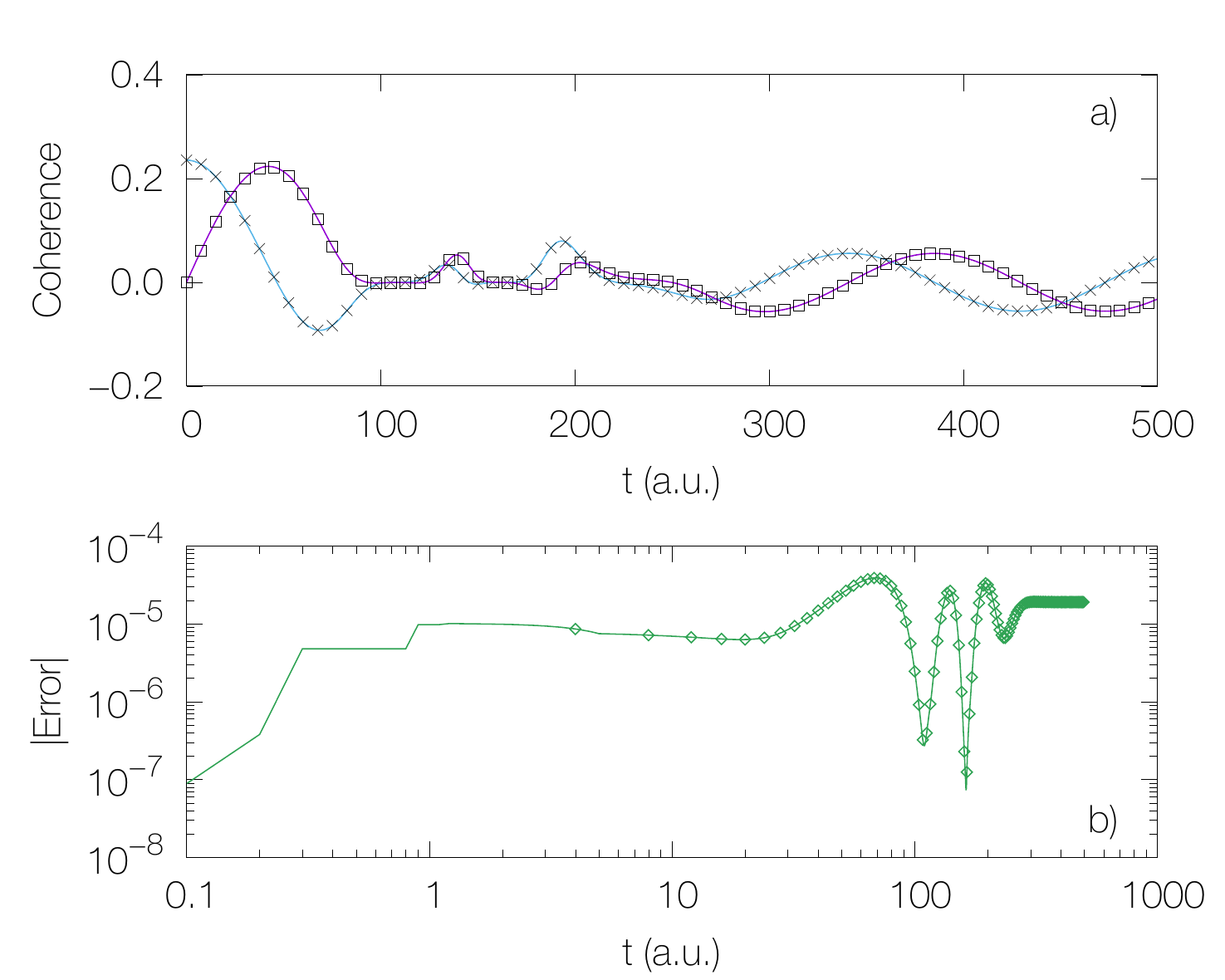}
\caption{a) Time evolution of the coherence $c_0(t)c_1^*(t)$ between the reference and the lowest
excited state: real part (sky blue line, TDUCC; black crosses, FCI) and imaginary part (dark violet
line, TDUCC; black open squares, FCI). b) Absolute deviation of the coherence from the exact
reference, on logarithmic axes, for the finest grid ($N_t = 5000$ steps over $t \in [0, 500]$ a.u.).
}
\label{fig:stacked_coherence}
\end{figure}

Figure~\ref{fig:stacked_coherence} shows the time evolution of the state coherence $c_0(t)c_1^*(t)$.
Over the considered time window the coherence departs from the exact result by at most $4.4 \times 10^{-5}$
against an amplitude of $0.24$, and the excited- and ground-state populations by at most $2 \times
10^{-5}$. Coherences require more work to track in non-Hermitian methods, where
imbalances between the left and right states introduce the mentioned asymmetrical
residues. Here the bra and ket come from the same norm-preserving unitary map, and the deviation in
the bottom panel oscillates about a slowly rising envelope. The
coherence error repeats the pattern of the dipole with a shallower initial slope. The
peak deviation falls from $8 \times 10^{-3}$ at $N_t = 100$ to $7 \times 10^{-4}$ at $N_t=250$, then
to $2 \times 10^{-4}$ at $N_t=500$ and $4 \times 10^{-5}$ at $N_t = 1000$, close to second-order
scaling, and finally stalls at $4 \times 10^{-5}$ when the grid is refined to $N_t = 5000$.

\begin{figure}[htp!]
\centering
\includegraphics[scale=0.55]{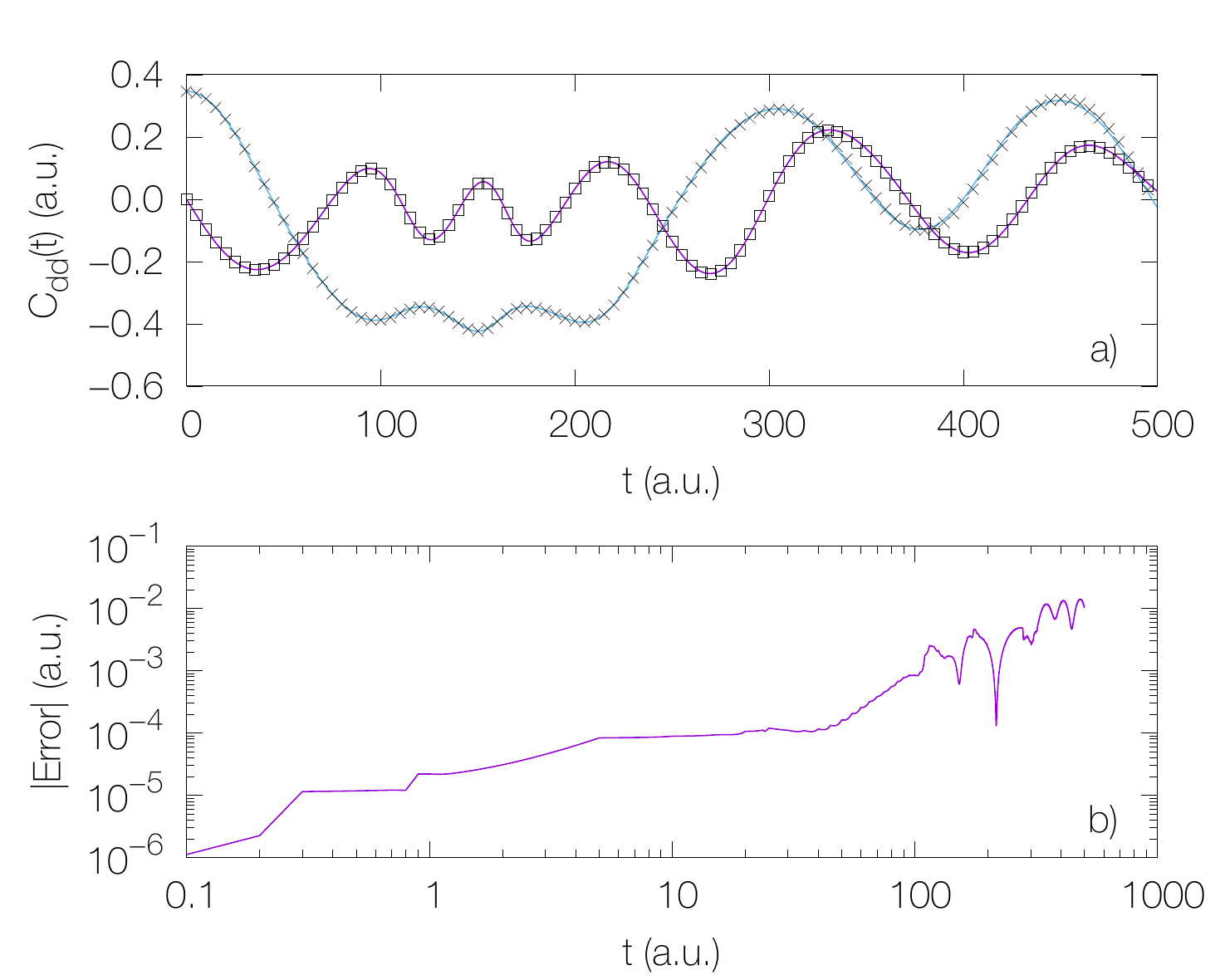}
\caption{a) Time evolution of the dipole correlation function
$C_{\mr{dd}}(t) = \langle\bar{\Psi}(0)|\hat{\mu}(t)\hat{\mu}(0)|\bar{\Psi}(0)\rangle$; real part
(sky blue line, TDUCC; black crosses, FCI) and imaginary part (dark violet line, TDUCC; black open
squares, FCI). b) Absolute deviation
$|C^{\mr{UCC}}_{\mr{dd}}(t) - C^{\mr{FCI}}_{\mr{dd}}(t)|$ from the exact reference, on logarithmic axes, for the finest
grid ($N_t = 5000$ steps over $t \in [0, 500]$ a.u.).}
\label{fig:stacked_corr_func}
\end{figure}

The time evolution of the correlation function, defined as
$C_{\mr{dd}}(t)=\langle\bar{\Psi}(0)|\hat{\mu}(t)\hat{\mu}(0)|\bar{\Psi}(0)\rangle$ (where
$\hat{\mu}(t)=\hat{U}_{\mathrm{QM}}^{\dagger}(t)\hat{\mu}\hat{U}_{\mathrm{QM}}(t)$), is presented in
Fig.~\ref{fig:stacked_corr_func}. Two propagated states determine this quantity: i) the
left (bra) factor is the ordinary state evolved in this work, while ii) the right (ket)
factor is initialized by the dipole acting on the correlated state,
$\hat{\mu}\,e^{\hat{T}}|\bar{\Psi}(0)\rangle$, or
$\bar{\mu}|\bar{\Psi}(0)\rangle=e^{-\hat{T}}\hat{\mu}\,e^{\hat{T}}|\bar{\Psi}(0)\rangle$ as used
here, and then evolved forward. The present formalism is not exact for this quantity; we include the
figure to show that limitation. The interaction-picture generator $\hat{y}_{\mr{i}}(t)$ is built to
propagate the physical superposition state $[c_0 + \tilde{z}(0)]|0\rangle$, which the
``dipole-kicked'' ket $\hat{\mu}\,e^{\hat{T}}|\bar{\Psi}(0)\rangle$ is not.  We note that the
deviation stops responding to the time step. Refining the grid from $N_t = 250$ to $N_t = 5000$
moves the peak deviation from $1.2 \times 10^{-2}$ to $1.4 \times 10^{-2}$ a.u. 

\begin{figure}[htp!]
\centering
\includegraphics[scale=0.55]{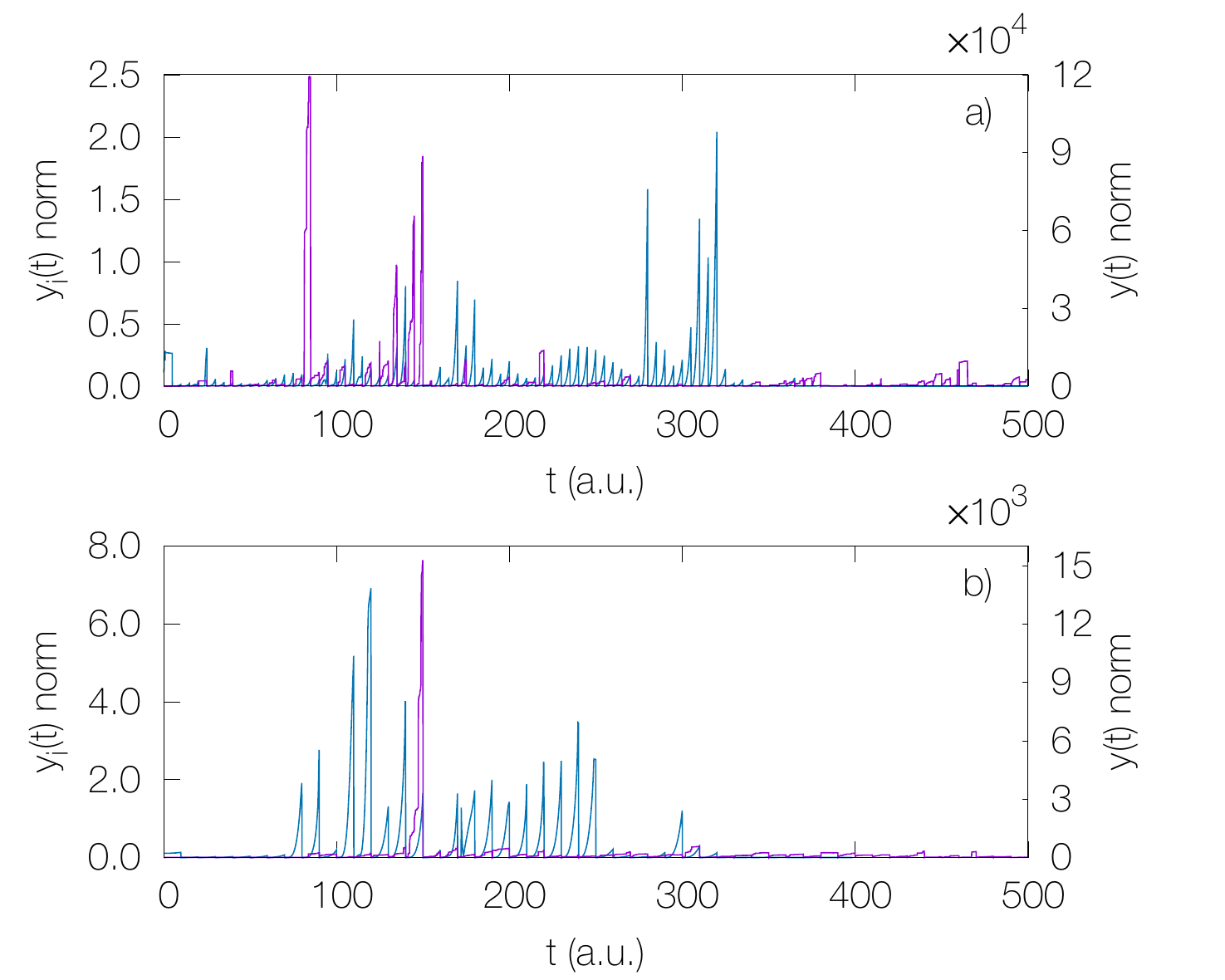}
\caption{Time-dependent cluster generator norms under the piecewise restart protocol of
Sec.~\ref{sec:piecewise}, for $N_t = 5000$ steps over $t \in [0, 500]$ a.u. Left-hand axis: the
interaction-picture generator norm $\|\hat{y}_{\mr{i}}(t)\|$ (blue), reset to zero at every restart
and therefore tracing a sawtooth; right-hand axis: $\|\hat{y}(t)\|$ (dark violet), rescaled by the
common factor annotated at the top right of each panel. a) $100$ restart epochs, where
$\|\hat{y}_{\mr{i}}\|$ stays below $2.1$; b) $50$ epochs, where it reaches $6.9$.}
\label{fig:td_norms}
\end{figure}

Figure~\ref{fig:td_norms} tracks the cluster generator norms and tests the piecewise
propagation protocol of Sec.~\ref{sec:piecewise}. As the external field drives the system away from
its original state, the continuous cluster amplitudes grow. Over this window $\|\hat{y}(t)\|$
reaches $1.2 \times 10^{5}$. Each restart absorbs the accumulated amplitudes into an updated
reference state and resets the generator to zero, so the interaction-picture norm
$\|\hat{y}_{\mr{i}}(t)\|$ traces the sawtooth seen in both panels. With $100$ time-intervals the
norm never exceeds $2.1$; with $50$ intervals it reaches $6.9$. Doubling the interval between
restarts costs a factor of three in the quantity the procedure controls. A small generator keeps the
BCH expansion in the regime where the truncated algebra could be a good approximation, which limits metric
degradation and keeps long propagations stable under strong driving. These results suggest that the
solution to the TD-UCC derived here should be adaptive. In other simulations with lower external
couplings, we noted that even one single interval for the whole propagation is sufficient,
reinforcing the idea that adaptation is required based on the operator norms. An improved adaptive
method does not necessarily mean that one has to work in the interaction picture; in preliminary
tests we also find cases where $\hat{y}(t)$ is numerically bounded in an appropriate manner, whereas
$\hat{y}_{\mr{i}}(t)$ is not (these will be investigated in the future).

\begin{figure}[htp!]
\centering
\includegraphics[scale=0.55]{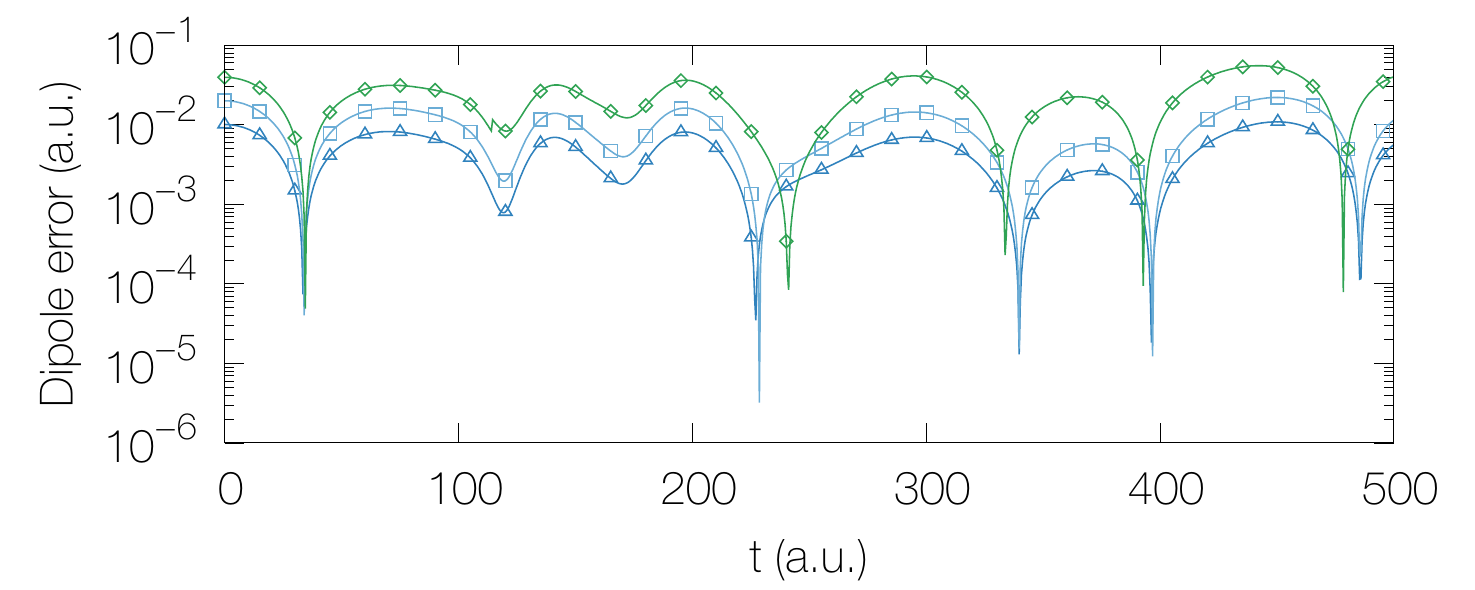}
\caption{TD dipole error, $|\mu^{\mr{UCC}}(t) - \mu^{\mr{FCI}}(t)|$, of the LC-initiated dipole, on
a logarithmic scale, for the detuned EOM-UCC problem, based on $\hat{T}'$ (this cluster operator is
used for the rest of the simulation after solving for the eigenvalues).  Time window $t \in [0,
500]$ a.u., 1000 time-steps. Results for $\lambda_{\mr{d}} = 0.75$ (medium blue, open triangles),
$\lambda_{\mr{d}}=0.5$ (light blue, open squares) and $\lambda_{\mr{d}}=0$ (green, open diamonds).}
\label{fig:detuned_performance}
\end{figure}

For the numerical cases above we find that
$\|\tilde{X}^{I\dagger}|0\rangle\|/\|\tilde{X}^{I}|0\rangle\| \approx 10^{-14}$: the annihilating
condition holds to machine precision. The term $\tilde{z}(0)|0\rangle$ reduces to its forward
component $\sum_I c_I \tilde{X}^I|0\rangle$. This is a property of the model considered here due to
the condition $\hat{b}^{\dagger}|1\rangle = 0$, and the type of excitation operator used here, which
is basically the boson creation operator that the Hamiltonian relies heavily on (in realistic
cases the Hamiltonian is more complex). The anti-Hermitian generator, however, still plays a
dominant role in the dynamics since its stationary condition and EOM-UCC equations lead to the
construction of the initial state. To obtain explicit contributions from
$\tilde{X}^{I\dagger}|0\rangle$, we detune the ground-state cluster operator as
$\hat{T}'=\lambda_{\mr{d}}\hat{T}$, where $\hat{T}$ is the correctly converged cluster operator used
before. The EOM-UCC problem is then solved using the $\hat{T}'$-transformed static Hamiltonian, and
the TD simulation is run with the resulting eigen-vectors/eigen-energies. For
$\lambda_{\mr{d}} = 0.75$ we observe (Fig.~\ref{fig:detuned_performance}) dipole errors on the
order of $10^{-3}$, which are then raised to around $10^{-2}$ at envelope level for
$\lambda_{\mr{d}}=0.5$. This error sustains itself for $\lambda_{\mr{d}}=0$. These show that the
construction of the initial state through the generator $\tilde{z}(0)$ is a viable pathway to
approximate observables. The norms $\|\tilde{X}^{I\dagger}|0\rangle\|/\|\tilde{X}^{I}|0\rangle\|$
here are $8\times 10^{-3}$, $2\times 10^{-2}$, and $3\times 10^{-2}$ for $\lambda_{\mr{d}} = 0.75$,
$\lambda_{\mr{d}} = 0.5$, and $\lambda_{\mr{d}} = 0$, respectively. In addition, we remark that
the object $\tilde{z}(0)$ exists as an anti-Hermitian operator and dynamics originator regardless of
detuning; this is due to its definition involving $\tilde{X}^{I\dagger}$ as a general operator.

\subsection{Theoretical}

In general, the conjugate vector $\tilde{X}^{I\dagger}$ enters the superposition generator as a consequence of the
anti-Hermitian restriction. Each $\hat{\kappa}_\mu$ combines an excitation $\hat{\tau}_{\bar{\mu}}$
with its de-excitation $\hat{\tau}_{\bar{\mu}}^\dagger$. Every forward vector carries its adjoint,
and through the nested commutators with $\hat{T}$ this adjoint does not annihilate the reference.
Because the $\hat{\Xi}_{0,\mu}$ are anti-Hermitian as well, $\tilde{X}^{I\dagger}$ lies in the same
excitation space as $\tilde{X}^I$. This pairing recalls the $X$/$Y$
amplitudes of RPA and TDHF, where de-excitations couple to correlations already present in the
ground state. The resemblance is also noted with the eigenvalue problem $\mathbf{A}\mathbf{X}^I =
\Omega_I \mathbf{S}\mathbf{X}^I$. The symmetric Jacobian together with the antisymmetric metric
$\langle[\hat{\Xi}_{0,\mu},\hat{\Xi}_{0,\nu}]\rangle$ yields a symplectic spectrum in $\pm\Omega_I$
pairs, with $\tilde{X}^I$ and $\tilde{X}^{I\dagger}$ acting as the forward/backward partners collected into
the static operator $\tilde{z}(0)=\sum_{I>0}(c_I\tilde{X}^I
- c_I^*\tilde{X}^{I\dagger})$. The algebraic form of Eq.~(\ref{eq:practical_eigenvalue}) is known: it
shows a double commutator against an antisymmetric metric, which after the symmetrization noted
above is the equation-of-motion method of Rowe \cite{rowe1968equations} with
$\hat{H}\to\hat{H}_T$. The eigenvalue problem here, however, arises as the unperturbed limit of
the variational equation (\ref{eq:z1_eom}), forming the basis onto which to project,
$\hat{\Xi}_{0,\mu}=\hat{\pi}_T(\hat{\kappa}_\mu)$. 

For dynamics the above does not hold. 
The propagator $\hat{U}_y(t)$ is applied to the frozen reference, as opposed to a linear superposition of
$e^{\mp\ui\Omega_I t}$ terms, so no explicit $e^{+\ui\Omega_I t}\tilde{X}^{I\dagger}|0\rangle$
term is propagated. The eigenrelation
$[\hat{H}_T,\tilde{X}^I]=\Omega_I\tilde{X}^I$, the operator form of
Eq.~(\ref{eq:practical_eigenvalue}), is applied to the correlated reference and is exact only
in the closed-algebra cases or FCI limits, whereas RPA usually restricts its eigenvalue problem to single
excitations based on an uncorrelated stationary state. The metric
$S_{\mu\nu}'(t)=\langle\bar{\Psi}(0)|\hat{\Xi}_{\mr{I},\mu}(t)\hat{\Xi}_{\mr{I},\nu}(t)|\bar{\Psi}(0)\rangle$
also follows the propagated basis in time, while the RPA metric is computed once over the frozen
reference and stays fixed. We keep the non-annihilating adjoint, since it arises from the anti-Hermitian
construction itself, and can be used to investigate linkedness and algebra closedness. If the
closure involves the Hamiltonian itself, it can be quite beneficial for size-extensivity,
even from the configuration interaction perspective.

Because $\tilde{X}^{I\dagger}|0\rangle\neq 0$ in general for truncated/approximated cases, the
``vacuum variance'' $\langle-\tilde{z}^2(0)\rangle$ is a required term. A direct expansion of
Eq.~(\ref{eq:heisenberg_b}) involves the product
$\langle\tilde{z}^2(0)\rangle\langle\tilde{B}_y(t)\rangle$, which can be cancelled out.  Given the
property $\tilde{z}^\dagger=-\tilde{z}$, the quadratic contribution is
$\tilde{z}(0)\tilde{B}_y(t)\tilde{z}(0)$, the same operator on both sides, and it splits into a
double commutator and an anticommutator; the form $\tilde{z}^\dagger\tilde{B}_y\tilde{z}$ would not.
The commutator $\langle[[\tilde{B}_y(t),\tilde{z}(0)],\tilde{z}(0)]\rangle$ is connected. In the
anticommutator, the fully disconnected part cancels against the normalization denominator
$|c_0|^2+\langle-\tilde{z}^2(0)\rangle$, and any term containing an isolated $\tilde{z}(0)$ factor
vanishes because of the requirement $\langle\tilde{z}(0)\rangle=0$.  The surviving contributions are
therefore connected, $\langle\{\tilde{B}_y(t),\tilde{z}^2(0)\}\rangle_{\mathrm{C}}$. Over the
reference $e^{\hat{T}}|0\rangle$ they reduce, through generalized Wick contraction
\cite{kutzelnigg1997normal, mukherjee1997normal}, to cumulants of the reference density matrices, so
Eq.~(\ref{eq:size_extensive_heisenberg}) is linked in the many-body sense, where closed ``vacuum''
factors drop out and open disconnected cumulants are kept. The cumulants are well-behaved as long as
they feature the correct locality.

Truncating the excitation manifold has consequences of algebraic and numerical character.  A plain
truncation comes at a cost in closure. In general, the unitary Hausdorff series does not
terminate, so a finite excitation space cannot close under it, in contrast to the terminating
expansion of standard similarity-transformed CC. The commutators
$[\hat{\Xi}_{0,\mu},\hat{\Xi}_{0,\nu}]$ can create outside terms. The similarity transform
$\tilde{B}_y(t)=e^{-\hat{y}}\hat{B}e^{\hat{y}}$ can leak out of the manifold and let the surviving
cumulants pick up nonlocal contributions; and the Jacobi argument for the symmetric Jacobian fails,
so $\mb{A}$ would have to be symmetrized by hand.

Closing only the $\mathrm{ad}_{\hat{y}}$ component of $\hat{B}$ makes $\tilde{B}_y$ well-behaved
while the trajectory $\hat{y}(t)$ stays approximate in nature. For a target dimension $M$,
completing the manifold to a closed $\mathfrak{su}(M)$ puts the propagation on the invariant
subspace $\mathcal{H}_M$ that the algebra generates, leaving the Hamiltonian coupling between
$\mathcal{H}_M$ and its complement. The FCI limit (dimension $N$) is the maximal case
$\mathcal{H}_N=\mathcal{H}$, where this coupling vanishes.  A closed $\mathfrak{su}(M)$ need not
be local, so additivity for separated subsystems requires the completion to feature locality;
otherwise small nonlocality leaks remain, as in truncated EOM-CC and ic-MRCC. The piecewise
restarts bound amplitude growth without restoring closure. In addition, applications will have to
take supermolecular additivity into account.

The Dirac-Frenkel strategy followed in this work is shared with the multiconfiguration
time-dependent Hartree method, tensor-network TDVP schemes, and variational quantum
simulation \cite{li2017efficient, yuan2019theory, mcardle2019variational}. Our working equation,
$\sum_\nu \mr{Im}[S_{\mu\nu}'(t)]\dot{y}_{\mr{i},\nu} = -\mr{Re}[h_{\mr{eff},\mu}(t)]$, has the
form of a metric acting on ``rates''. Some differences: the
real/imaginary split follows from the anti-Hermitian generator, while the metric and the
$\hat{\Xi}_{\mr{I},\mu}$ operators come from the internally contracted cluster algebra after SVD
orthogonalization and QR rotation. Both could serve as a starting point for further work on unitary
TD theories and their numerics.

\section{Conclusion}

We have presented a time-dependent unitary coupled-cluster approach to the propagation of general
quantum states. A time-dependent anti-Hermitian cluster operator generates the non-ground-state wave
function component, while the evolution is unitary, and norms are conserved.  The observables are
strictly real-valued. Variations of the Dirac-Frenkel action give equations of motion expressed
through Heisenberg-like commutators. The unperturbed dynamics yields the EOM-UCC eigenvalue problem,
which is based on a symmetric Jacobian.  For the free propagation of general initial states, the
observable expressions can be linked, with the unphysical vacuum factor cancelling against the
normalization. Size-extensivity, however, must be considered in detail as EOM methods are still
subject to unwanted non-localities. The numerical tests reproduced the reference dipoles,
populations, and coherences, while the piecewise restarts kept long propagations stable. The
correlation function is not exact in the present formulation; treating it, and applying the theory
to realistic atomic and molecular systems, are potential future steps.

\funding{This material is based upon work supported by the U.S. Department of Energy, Office of
Science, Office of Basic Energy Sciences, Early Career Research Program, Award Number
DE-SC-0025662.}

\roles{Mart\'in A. Mosquera: Conceptualization, Methodology, Software, Formal analysis,
Investigation, Writing -- original draft, Writing -- review and editing.}

\data{The data that support the findings of this study are openly available at the following
URL/DOI: \url{https://doi.org/10.5281/zenodo.22017136}. The deposit contains the simulation code,
the input files, and the propagation and eigenvalue output underlying every figure and table in this
article and in the supplementary material. It is released under a CC BY 4.0 licence.}

\suppdata{The supplementary material is appended to this preprint (Secs.~S1--S5). Derivation of the Maurer-Cartan relation and the
first-order generator equation of motion, linked observable expressions, observable matrix elements
and their supermolecular additivity, and results for truncated excitation spaces including the
active space and CAS-CI eigenvectors, ground-state and excitation energies, converged cluster
amplitudes, and the first EOM-UCC excitation vector.}

\bibliographystyle{iopart-num}
\bibliography{refs}

\clearpage

\setlength{\heavyrulewidth}{\lightrulewidth}
\setlength{\cmidrulewidth}{\lightrulewidth}

\setcounter{section}{0}
\setcounter{figure}{0}
\setcounter{table}{0}
\setcounter{equation}{0}
\renewcommand{\thesection}{S\arabic{section}}
\renewcommand{\thesubsection}{S\arabic{section}.\arabic{subsection}}
\renewcommand{\thefigure}{S\arabic{figure}}
\renewcommand{\thetable}{S\arabic{table}}
\renewcommand{\theequation}{S\arabic{equation}}
\renewcommand{\theHsection}{S\arabic{section}}
\renewcommand{\theHsubsection}{S\arabic{section}.\arabic{subsection}}
\renewcommand{\theHfigure}{S\arabic{figure}}
\renewcommand{\theHtable}{S\arabic{table}}
\renewcommand{\theHequation}{S\arabic{equation}}

{\fontsize{18}{21}\selectfont\noindent\raggedright\textsf{Supplementary material:
Observables and anti-Hermitian generators in time-dependent unitary coupled cluster
theory}\par}

\vspace{5mm}
\noindent{\fontsize{10}{12}\selectfont Mart\'in A. Mosquera}

\vspace{2mm}
\noindent{\fontsize{8}{10}\selectfont\itshape Department of Chemistry and Biochemistry,
Montana State University, Bozeman, Montana 59717, USA}

\vspace{6mm}

\section{Maurer-Cartan Relation}\label{sec:appendix_a}

Here we derive the variations $\hat{\Xi}_\mu^{(n)}(t)$ and the first-order residual
$\hat{R}^{(1)}(t)$ used in the main text. Both follow from derivatives of the exponential map
$\hat{U}_{\mr{m}} = \exp[\hat{x}_{\mr{m}}(t; g)]$ in the continuous parameter $g$, evaluated at
$g=0$.

The residual is $\hat{R}_{\mr{m}}(t; g) = \ui\,\hat{\pi}_{x_{\mr{m}}}(\partial_t \hat{x}_{\mr{m}}) -
\hat{H}_{x_{\mr{m}}}$, with the transformed Hamiltonian $\hat{H}_{x_{\mr{m}}} =
e^{-\hat{x}_{\mr{m}}}\hat{H}e^{\hat{x}_{\mr{m}}}$ and the transformed basis operator
$\hat{\Xi}_{\mr{m},\mu}(t; g) = \hat{\pi}_{x_{\mr{m}}}(\hat{\kappa}_\mu)$. The continuous generator
is $\tilde{z}(t; g) = \hat{\pi}_{x_{\mr{m}}}(\partial_g \hat{x}_{\mr{m}})$. We use the Maurer-Cartan
relation for the projection map [Eq.~(3) of Ref.~\cite{mosquera2025}]:
\begin{equation}
    \partial_\alpha (\hat{\pi}_u (\partial_\beta \hat{u})) - \partial_\beta (\hat{\pi}_u (\partial_\alpha \hat{u})) = [ \hat{\pi}_u (\partial_\beta \hat{u}), \hat{\pi}_u (\partial_\alpha \hat{u}) ]~.
\end{equation}
Setting $\hat{u} = \hat{x}_{\mr{m}}$, $\beta = g$, and $\alpha = x_{\mr{m},\mu}$ gives
$\hat{\pi}_u(\partial_\beta \hat{u}) = \tilde{z}(t; g)$ and $\hat{\pi}_u(\partial_\alpha \hat{u}) =
\hat{\Xi}_{\mr{m},\mu}$, so
\begin{equation}
    \partial_g \hat{\Xi}_{\mr{m}, \mu} = \partial_\mu \tilde{z}(t; g) + [\hat{\Xi}_{\mr{m}, \mu}, \tilde{z}(t; g)]~.
\end{equation}
Evaluated at $g=0$, this yields the first-order variation
\begin{equation}
    \hat{\Xi}_\mu^{(1)}(t) = \partial_\mu \tilde{z}(t) + [\hat{\Xi}_\mu^{(0)}(t), \tilde{z}(t)]~,
    \qquad \hat{\Xi}_\mu^{(0)}(t) = \hat{\pi}_x(\hat{\kappa}_\mu)~,
\end{equation}
in agreement with the main text.

Taking instead $\alpha = t$ and $\beta = g$ gives
\begin{equation}
    \partial_g \left[ \hat{\pi}_{x_{\mr{m}}}(\partial_t \hat{x}_{\mr{m}}) \right] = \partial_t \tilde{z}(t; g) + [\hat{\pi}_{x_{\mr{m}}}(\partial_t \hat{x}_{\mr{m}}), \tilde{z}(t; g)]~,
\end{equation}
and, since the transformed Hamiltonian is a similarity transform of a fixed operator, its derivative
is the commutator
\begin{equation}
    \partial_g \hat{H}_{x_{\mr{m}}} = [\hat{H}_{x_{\mr{m}}}, \tilde{z}(t; g)]~.
\end{equation}
Combining the last two relations, the derivative of the residual is
\begin{equation}
    \partial_g \hat{R}_{\mr{m}} = \ui\, \partial_g \left[ \hat{\pi}_{x_{\mr{m}}}(\partial_t \hat{x}_{\mr{m}}) \right] - \partial_g \hat{H}_{x_{\mr{m}}} = \ui\, \partial_t \tilde{z}(t; g) + [\hat{R}_{\mr{m}}, \tilde{z}(t; g)]~.
\end{equation}
Evaluated at $g=0$, with $\tilde{z}(t) = \tilde{z}(t; g)\big|_{g=0}$ and $\hat{R}_x =
\hat{R}_{\mr{m}}\big|_{g=0}$, this gives
\begin{equation}
    \hat{R}^{(1)}(t) = \ui\, \partial_t \tilde{z}(t) + [\hat{R}_x(t), \tilde{z}(t)]~,
\end{equation}
which is the first-order residual quoted in the main text.

\section{First-Order Generator Equation of Motion}\label{sec:first_order_eom}

The generator $\tilde{z}(t)$ obeys an equation of motion obtained by differentiating the propagation
condition
\begin{equation}
    \langle [ \hat{\Xi}_{\mr{m},\mu}(t; g) \, , \, \hat{R}_{\mr{m}}(t; g) ] \rangle = 0
\end{equation}
with respect to $g$ and evaluating at $g=0$. With $\hat{\Xi}_{\mr{m},\mu}\big|_{g=0} =
\hat{\Xi}_\mu^{(0)}(t)$, $\partial_g\hat{\Xi}_{\mr{m},\mu}\big|_{g=0} = \hat{\Xi}_\mu^{(1)}(t)$,
$\hat{R}_{\mr{m}}\big|_{g=0} = \hat{R}_x(t)$, and $\partial_g\hat{R}_{\mr{m}}\big|_{g=0} =
\hat{R}^{(1)}(t)$, this gives
\begin{equation}
    \langle [ \hat{\Xi}_\mu^{(1)}(t) \, , \, \hat{R}_x(t) ] \rangle + \langle [ \hat{\Xi}_\mu^{(0)}(t) \, , \, \hat{R}^{(1)}(t) ] \rangle = 0~.
\end{equation}
Substituting the first-order residual $\hat{R}^{(1)}(t) = \ui\partial_t\tilde{z}(t) + [\hat{R}_x(t),
\tilde{z}(t)]$ from Sec.~\ref{sec:appendix_a} and isolating the time derivative,
\begin{equation}
    \langle [ \hat{\Xi}_\mu^{(0)}(t) \, , \, \ui\partial_t\tilde{z}(t) ] \rangle = -\langle [ \hat{\Xi}_\mu^{(0)}(t) \, , \, [\hat{R}_x(t), \tilde{z}(t)] ] \rangle - \langle [ \hat{\Xi}_\mu^{(1)}(t) \, , \, \hat{R}_x(t) ] \rangle~.
\end{equation}
This is the working equation for the first-order generator shown in the main text. It applies to
both free and driven propagation and is accurate at short times. The reference operator
$\hat{x}(t)$ carries the nonlinear part of the evolution, while $\tilde{z}(t)$ tracks the linear
response.

\section{Linked Observable Expressions}\label{sec:computation_observables}

Let the reference be $|\bar{\Psi}(0)\rangle = [c_0 +
\tilde{z}(0)]|0\rangle$, with $\tilde{z}(0)$ anti-Hermitian ($\tilde{z}^\dagger(0) =
-\tilde{z}(0)$), and let $\tilde{B}_y(t) = \hat{U}_y^\dagger(t)\hat{B}\hat{U}_y(t)$. 
The expectation value is
\begin{equation}\label{eq:si_obs_start}
    \langle \hat{B} \rangle(t) = \frac{ \langle [c_0^* - \tilde{z}(0)] \tilde{B}_y(t) [c_0 + \tilde{z}(0)] \rangle }{ |c_0|^2 + \langle -\tilde{z}^2(0) \rangle }~.
\end{equation}
The denominator $D \equiv |c_0|^2 + \langle -\tilde{z}^2(0) \rangle$ is the squared norm $\langle
\bar{\Psi}(0)|\bar{\Psi}(0)\rangle$; its linear cross terms vanish because $\langle \tilde{z}(0)
\rangle = 0$, enforced by the reference-component convention adopted in the main text. Each
eigenvector satisfies $\langle \tilde{X}^I\rangle=0$, 
which each commutator of the working equations ignores. 

Expanding the numerator gives
\begin{multline}
    \langle [c_0^* - \tilde{z}(0)] \tilde{B}_y(t) [c_0 + \tilde{z}(0)] \rangle
    = |c_0|^2 \langle \tilde{B}_y(t) \rangle + c_0^* \langle \tilde{B}_y(t) \tilde{z}(0) \rangle \\
    - c_0 \langle \tilde{z}(0) \tilde{B}_y(t) \rangle - \langle \tilde{z}(0) \tilde{B}_y(t) \tilde{z}(0) \rangle~.
\end{multline}
The quadratic term has the same operator on both sides, so it splits into a double
commutator and an anticommutator,
\begin{equation}
    -\langle \tilde{z}(0) \tilde{B}_y(t) \tilde{z}(0) \rangle = \frac{1}{2} \langle [[\tilde{B}_y(t), \tilde{z}(0)], \tilde{z}(0)] \rangle - \frac{1}{2} \langle \{ \tilde{B}_y(t), \tilde{z}^2(0) \} \rangle~.
\end{equation}
Wick's theorem separates the anticommutator into disconnected and connected parts,
\begin{equation}
    -\frac{1}{2} \langle \{ \tilde{B}_y(t), \tilde{z}^2(0) \} \rangle = -\langle \tilde{B}_y(t) \rangle \langle \tilde{z}^2(0) \rangle - \frac{1}{2} \langle \{ \tilde{B}_y(t), \tilde{z}^2(0) \} \rangle_{\mr{C}}~,
\end{equation}
where the subscript $\mr{C}$ denotes the connected contribution. The disconnected piece combines
with the leading term,
\begin{equation}
    |c_0|^2 \langle \tilde{B}_y(t) \rangle - \langle \tilde{B}_y(t) \rangle \langle \tilde{z}^2(0) \rangle = \langle \tilde{B}_y(t) \rangle \left( |c_0|^2 - \langle \tilde{z}^2(0) \rangle \right) = D \langle \tilde{B}_y(t) \rangle~,
\end{equation}
so dividing by $D$ removes it and leaves $\langle \tilde{B}_y(t) \rangle$. The observable becomes
\begin{multline}\label{eq:si_size_extensive}
    \langle \hat{B} \rangle(t) = \langle \tilde{B}_y(t) \rangle
    + \frac{1}{D} \Big( c_0^* \langle \tilde{B}_y(t) \tilde{z}(0) \rangle
    - c_0 \langle \tilde{z}(0) \tilde{B}_y(t) \rangle \\
    + \frac{1}{2} \langle [[\tilde{B}_y(t), \tilde{z}(0)], \tilde{z}(0)] \rangle
    - \frac{1}{2} \langle \{ \tilde{B}_y(t), \tilde{z}^2(0) \} \rangle_{\mr{C}} \Big)~,
\end{multline}
which is the size-extensive form used in the main text. The disconnected vacuum factor, the
product of $\langle \tilde{B}_y(t) \rangle$ and $\langle \tilde{z}^2(0) \rangle$, has cancelled
against the normalization.

\section{Observable Matrix Elements}\label{sec:transition_moments}

The observable of Sec.~\ref{sec:computation_observables} is quadratic in $\tilde{z}(0)$, where
initial conditions can reveal matrix elements between EOM-UCC states in the
same linked form, with no additional requirements. We derive the expressions here.

The relevant operator is the transformed object:
\begin{equation}\label{eq:si_BT_def}
    \tilde{B}_T \equiv e^{-\hat{T}} \hat{B} e^{\hat{T}}~.
\end{equation}
The correlated states are $|\Psi_K\rangle = e^{\hat{T}}\tilde{X}^K|0\rangle$ ($|\Psi_0\rangle =
e^{\hat{T}}|0\rangle$).
Since $\hat{T}$ is anti-Hermitian, $e^{\hat{T}\dagger} = e^{-\hat{T}}$, and we have that
$\langle \Psi_I | \hat{B} | \Psi_J \rangle = \langle \Phi_I | \tilde{B}_T | \Phi_J
\rangle$ with $|\Phi_K\rangle = \tilde{X}^K|0\rangle$.

We denote $\langle \cdot \rangle \equiv \langle 0 | \cdot | 0 \rangle$, and assume the condition
$\tilde{X}^{K\dagger}|0\rangle = 0$, equivalently $\langle 0 |\tilde{X}^K = 0$. This condition holds
in the FCI limit or closed-algebra cases. Taking the adjoint of the eigenrelation $[\hat{H}_T, \tilde{X}^K] = \Omega_K
\tilde{X}^K$ gives $[\hat{H}_T, \tilde{X}^{K\dagger}] = -\Omega_K \tilde{X}^{K\dagger}$.
Applying this to the reference, with $\hat{H}_T|0\rangle = E_0|0\rangle$ in the same limit, yields
\begin{equation}
    \hat{H}_T \left( \tilde{X}^{K\dagger}|0\rangle \right) = (E_0 - \Omega_K) \left( \tilde{X}^{K\dagger}|0\rangle \right)~,
\end{equation}
which demands that
$\tilde{X}^{K\dagger}|0\rangle = 0$. In general, under truncations/approximations the relations derived next, and the
annihilating condition, hold only approximately, but these become exact in the FCI limit, or
closed-algebraic systems. We now
write
\begin{equation}\label{eq:si_deltaB}
    \delta \tilde{B}_T \equiv \tilde{B}_T - \langle \tilde{B}_T \rangle
\end{equation}
for the fluctuation operator, and $S_{IJ} \equiv \langle \Phi_I | \Phi_J \rangle$ for the overlap
between eigenstates.

The initial generator $\tilde{z}(0) = \sum_{I>0} ( c_I \tilde{X}^I - c_I^* \tilde{X}^{I\dagger} )$
admits complex coefficients. The two phase choices $c_K = 1$ and $c_K = \ui$ produce the
anti-Hermitian pair
\begin{equation}\label{eq:si_zw_def}
    z_K = \tilde{X}^K - \tilde{X}^{K\dagger}~, \qquad
    w_K = \ui \left( \tilde{X}^K + \tilde{X}^{K\dagger} \right)~,
\end{equation}
whose action on the reference follows from the annihilating condition,
\begin{equation}\label{eq:si_zw_action}
    z_K|0\rangle = |\Phi_K\rangle~,\quad \langle 0 | z_K = -\langle \Phi_K |~, \qquad
    w_K|0\rangle = \ui|\Phi_K\rangle~,\quad \langle 0 | w_K = +\ui\langle \Phi_K |~.
\end{equation}
The relative sign between bra and ket is negative for $z_K$ and positive for $w_K$. This single
asymmetry is what separates the real and imaginary parts below.

Let $a$ and $b$ be any two anti-Hermitian operators. Expanding both orderings of the double
commutator, we get:
\begin{align}
    \langle [[\tilde{B}_T,a],b] \rangle &= \langle \tilde{B}_T ab \rangle - \langle a \tilde{B}_T b \rangle - \langle b \tilde{B}_T a \rangle + \langle ba \tilde{B}_T \rangle~, \\
    \langle [[\tilde{B}_T,b],a] \rangle &= \langle \tilde{B}_T ba \rangle - \langle b \tilde{B}_T a \rangle - \langle a \tilde{B}_T b \rangle + \langle ab \tilde{B}_T \rangle~.
\end{align}
We can also note that
\begin{equation}\label{eq:si_engine}
    \langle [[\tilde{B}_T,a],b] \rangle + \langle [[\tilde{B}_T,b],a] \rangle = \langle \{ \tilde{B}_T, \{a,b\} \} \rangle
    - 2 \left( \langle a \tilde{B}_T b \rangle + \langle b \tilde{B}_T a \rangle \right)~.
\end{equation}
The anticommutator above is not linked. It contains the disconnected piece
$2\langle \tilde{B}_T \rangle\langle\{a,b\}\rangle$. We therefore define the bilinear functional with its
connected part, in the same convention as
Sec.~\ref{sec:computation_observables},
\begin{equation}\label{eq:si_F_def}
\begin{split}
    F_{\mr{C}}[a,b] &\equiv \frac{1}{2}\left( \langle [[\tilde{B}_T,a],b] \rangle + \langle [[\tilde{B}_T,b],a] \rangle \right)
    - \frac{1}{2} \langle \{ \tilde{B}_T, \{a,b\} \} \rangle_{\mr{C}}~, \\
    \langle \{ \tilde{B}_T, W \} \rangle_{\mr{C}} &= \langle \{ \tilde{B}_T, W \} \rangle - 2 \langle \tilde{B}_T \rangle \langle W \rangle~.
\end{split}
\end{equation}
Combining Eqs.~(\ref{eq:si_engine}) and (\ref{eq:si_F_def}) gives the compact evaluation
\begin{equation}\label{eq:si_F_value}
    F_{\mr{C}}[a,b] = -\left( \langle a\, \delta \tilde{B}_T\, b \rangle + \langle b\, \delta \tilde{B}_T\, a \rangle \right)~,
\end{equation}
that is, the plain result with $\tilde{B}_T$ replaced by the fluctuation operator of
Eq.~(\ref{eq:si_deltaB}). This form makes the linkedness explicit. Since $\langle a \rangle =
\langle b \rangle = 0$, the generalized Wick contraction over the correlated reference, discussed in
the main text, reduces $\langle a\, \delta \tilde{B}_T\, b \rangle$. 
The invariance of all terms of Eq.~(\ref{eq:si_F_def}) under $\hat{B} \to \hat{B} + c$
is a check of this reduction, but not a proof of linkedness.
Setting $a = b = \tilde{z}(0)$ recovers the quadratic term of
Sec.~\ref{sec:computation_observables}: $F_{\mr{C}}[\tilde{z}(0),\tilde{z}(0)] = -2\langle
\tilde{z}(0)\, \delta \tilde{B}_T\, \tilde{z}(0)\rangle$, which is twice the combination that
Eq.~(\ref{eq:si_size_extensive}) is left with once the disconnected vacuum factor has cancelled
against the normalization.

\subsection{Real and imaginary parts}

Taking $a = z_I$ and $b = z_J$ in Eq.~(\ref{eq:si_F_value}) and inserting
Eq.~(\ref{eq:si_zw_action}), the two sign flips multiply to $+1$,
\begin{equation}\label{eq:si_F_real}
    F_{\mr{C}}[z_I,z_J] = \langle \Phi_I | \delta \tilde{B}_T | \Phi_J \rangle + \langle \Phi_J | \delta \tilde{B}_T | \Phi_I \rangle
    = 2\,\mr{Re}\,\langle \Phi_I | \delta \tilde{B}_T | \Phi_J \rangle~,
\end{equation}
the last step using the Hermiticity of $\hat{B}$. Taking instead $a = z_I$ and $b = w_J$, only one
sign flips, and the two terms subtract rather than add,
\begin{equation}\label{eq:si_F_imag}
    F_{\mr{C}}[z_I,w_J] = \ui \left( \langle \Phi_I | \delta \tilde{B}_T | \Phi_J \rangle - \langle \Phi_J | \delta \tilde{B}_T | \Phi_I \rangle \right)
    = -2\,\mr{Im}\,\langle \Phi_I | \delta \tilde{B}_T | \Phi_J \rangle~.
\end{equation}
Combining the two last results gives the complex transition moment in linked form,
\begin{equation}\label{eq:si_transition_moment}
    \;\langle \Phi_I | \tilde{B}_T | \Phi_J \rangle - \langle \tilde{B}_T \rangle S_{IJ}
    = \frac{1}{2} F_{\mr{C}}[z_I,z_J] - \frac{\ui}{2} F_{\mr{C}}[z_I,w_J]\;
\end{equation}
Every term on the right is a commutator or a connected anticommutator of $\tilde{B}_T$ with the
static EOM-UCC vectors, so the size-extensivity argument of
Sec.~\ref{sec:computation_observables} carries over.

Under $\hat{B} \to \hat{B} + c$ the right-hand side is invariant, whereas $\langle \Phi_I |
\tilde{B}_T | \Phi_J \rangle \to \langle \Phi_I | \tilde{B}_T | \Phi_J \rangle + c\,S_{IJ}$. A
transition moment between non-orthogonal states carries a disconnected piece $\langle \tilde{B}_T
\rangle S_{IJ}$, which scales with system size because $\langle \tilde{B}_T \rangle$ does.
Eq.~(\ref{eq:si_transition_moment}) isolates the linked remainder. It is possible to assume that
$S_{IJ} = \delta_{IJ}$ since this will hold true in the FCI limit, the subtraction vanishes for $I
\neq J$, and the right-hand side is the transition matrix element itself,
\begin{equation}\label{eq:si_transition_orthogonal}
    \langle \Phi_I | \tilde{B}_T | \Phi_J \rangle
    = \frac{1}{2} F_{\mr{C}}[z_I,z_J] - \frac{\ui}{2} F_{\mr{C}}[z_I,w_J]~,
    \qquad I \neq J~.
\end{equation}
Two special cases follow from Eq.~(\ref{eq:si_transition_moment}) with no further work.

We now analyze ground-to-excited state elements. The reference $|\Phi_0\rangle = |0\rangle$ is not
generated by an EOM vector. In Eq.~(\ref{eq:si_obs_start}) it enters through the amplitude $c_0$ rather. 
The corresponding elements are thus obtained 
from Eq.~(\ref{eq:si_zw_action}) with a single commutator. Using
$z_I|0\rangle = |\Phi_I\rangle$ and $\langle 0 | z_I = -\langle \Phi_I |$,
\begin{equation}\label{eq:si_ref_real}
    \langle [ \tilde{B}_T, z_I ] \rangle = \langle 0 | \tilde{B}_T | \Phi_I \rangle + \langle \Phi_I | \tilde{B}_T | 0 \rangle
    = 2\,\mr{Re}\,\langle \Phi_I | \tilde{B}_T | \Phi_0 \rangle~,
\end{equation}
while the conjugate $w_I$, for which both signs are positive, gives
\begin{equation}\label{eq:si_ref_imag}
    \langle [ \tilde{B}_T, w_I ] \rangle = \ui \left( \langle 0 | \tilde{B}_T | \Phi_I \rangle - \langle \Phi_I | \tilde{B}_T | 0 \rangle \right)
    = 2\,\mr{Im}\,\langle \Phi_I | \tilde{B}_T | \Phi_0 \rangle~.
\end{equation}
Hence
\begin{equation}\label{eq:si_ref_moment}
    \langle \Phi_I | \tilde{B}_T | \Phi_0 \rangle
    = \frac{1}{2} \langle [ \tilde{B}_T, z_I ] \rangle + \frac{\ui}{2} \langle [ \tilde{B}_T, w_I ] \rangle~.
\end{equation}
And we also identify:
\begin{equation}\label{eq:si_ref_linear_terms}
    c_0^* \langle \tilde{B}_T \tilde{z}(0) \rangle - c_0 \langle \tilde{z}(0) \tilde{B}_T \rangle
    = 2\,\mr{Re} \left( c_0 \langle \Phi_I | \tilde{B}_T | \Phi_0 \rangle \right)~,
\end{equation}
so that $c_0 = 1$ and $c_0 = \ui$ again select the real and imaginary parts.

We set $J = I$ in Eq.~(\ref{eq:si_transition_moment}) and
$S_{II} = 1$, and consider
Eq.~(\ref{eq:si_F_value}),
\begin{equation}\label{eq:si_diag_imag}
    F_{\mr{C}}[z_I,w_I] = -\left( \langle z_I\, \delta \tilde{B}_T\, w_I \rangle + \langle w_I\, \delta \tilde{B}_T\, z_I \rangle \right)
    = -\left( -\ui + \ui \right) \langle \Phi_I | \delta \tilde{B}_T | \Phi_I \rangle = 0~.
\end{equation}
In contrast, $F_{\mr{C}}[z_I,z_I] = \langle [[\tilde{B}_T,z_I],z_I] \rangle - \langle \{
\tilde{B}_T, z_I^2 \} \rangle_{\mr{C}}$ from Eq.~(\ref{eq:si_F_def}), so:
\begin{equation}\label{eq:si_diag_moment}
    \langle \Phi_I | \tilde{B}_T | \Phi_I \rangle = \langle \tilde{B}_T \rangle
    + \frac{1}{2} \langle [[ \tilde{B}_T, z_I ], z_I ] \rangle
    - \frac{1}{2} \langle \{ \tilde{B}_T, z_I^2 \} \rangle_{\mr{C}}~.
\end{equation}
This is precisely Eq.~(\ref{eq:si_size_extensive}) evaluated at $c_0 = 0$, $\tilde{z}(0) = z_I$ and
$D = \langle -z_I^2 \rangle = 1$, which closes the consistency of the two sections. The
excited-state expectation value is the superposition observable of
Sec.~\ref{sec:computation_observables} with the reference amplitude switched off. The above can be
used to estimate said element, which converges to the FCI limit.

\subsection{Supermolecular additivity}\label{sec:supermolecular}

We start from Eq.~(\ref{eq:si_obs_start}) at $t=0$, where $\hat{y}(0) = \hat{T}$ and
$\tilde{B}_y(0) = \tilde{B}_T$. Let $\mr{A}$ and $\mr{B}$ be non-interacting, $|0\rangle =
|0_{\mr{A}}\rangle \otimes |0_{\mr{B}}\rangle$, and $\hat{T} = \hat{T}_{\mr{A}} + \hat{T}_{\mr{B}}$
with $[\hat{T}_{\mr{A}},\hat{T}_{\mr{B}}] = 0$. For $\hat{B} = \hat{B}_{\mr{A}} + \hat{B}_{\mr{B}}$
the transformed operator splits,
\begin{equation}\label{eq:si_super_split}
    \tilde{B}_T = \tilde{B}_{\mr{A}} + \tilde{B}_{\mr{B}}~, \qquad
    \tilde{B}_{\mr{X}} = e^{-\hat{T}_{\mr{X}}} \hat{B}_{\mr{X}} e^{\hat{T}_{\mr{X}}}~,\qquad \mr{X}
    = \mr{A},~\mr{B}~,
\end{equation}
because $\hat{T}_{\mr{B}}$ commutes with $\hat{B}_{\mr{A}}$ and conversely. The same split holds for
$\tilde{B}_y(t)$ at later times as long as the Hamiltonian, the drive and the excitation space are
local, so that $\hat{y}(t)$ stays additive.

Let $z_{\mr{X}}$ be anti-Hermitian and local to $\mr{X}$, with $[z_{\mr{A}},z_{\mr{B}}] = 0$, and
write
\begin{equation}\label{eq:si_super_local}
    |\phi_{\mr{X}}\rangle = z_{\mr{X}}|0_{\mr{X}}\rangle~, \quad
    n_{\mr{X}} = \langle \phi_{\mr{X}} | \phi_{\mr{X}} \rangle
    = \langle -z_{\mr{X}}^2 \rangle~, \quad
    b_{\mr{X}} = \langle 0_{\mr{X}} | \tilde{B}_{\mr{X}} | 0_{\mr{X}} \rangle~, \quad
    t_{\mr{X}} = \langle 0_{\mr{X}} | \tilde{B}_{\mr{X}} | \phi_{\mr{X}} \rangle~,
\end{equation}
with $b_{\mr{X}}^{\mr{exc}} = \langle \phi_{\mr{X}} | \tilde{B}_{\mr{X}} | \phi_{\mr{X}} \rangle /
n_{\mr{X}}$; only the product $n_{\mr{X}} b_{\mr{X}}^{\mr{exc}}$ appears, so $n_{\mr{X}} \to 0$
is benign. Expectation values of operators on disjoint subsystems factorize. We also have that
$\langle z_{\mr{X}} \rangle = \langle 0_{\mr{X}} | \phi_{\mr{X}} \rangle = 0$ from the convention
$\tilde{X}^I \leftarrow \tilde{X}^I - \langle \tilde{X}^I \rangle \hat{1}$ of the main text; its
role is examined at the end.

Let us consider the case of one subsystem being excited. We take $\tilde{z}(0) = z_{\mr{A}} + z_{\mr{B}}$, for which
\begin{equation}\label{eq:si_super_components}
    [c_0 + \tilde{z}(0)]|0\rangle = c_0 |0\rangle + |\Phi_{\mr{A}}\rangle + |\Phi_{\mr{B}}\rangle~,
    \qquad |\Phi_{\mr{X}}\rangle = z_{\mr{X}}|0\rangle~,
\end{equation}
and three orthogonal components: $\langle 0 | \Phi_{\mr{X}} \rangle = \langle z_{\mr{X}} \rangle = 0$
and $\langle \Phi_{\mr{A}} | \Phi_{\mr{B}} \rangle = -\langle z_{\mr{A}} \rangle \langle z_{\mr{B}}
\rangle = 0$. The generators commute, so $\langle -\tilde{z}^2(0) \rangle = n_{\mr{A}} + n_{\mr{B}}
- 2 \langle z_{\mr{A}} \rangle \langle z_{\mr{B}} \rangle$, and the denominator is additive,
\begin{equation}\label{eq:si_super_norm}
    D = |c_0|^2 + n_{\mr{A}} + n_{\mr{B}}~.
\end{equation}
Each matrix element reduces to local scalars. These are $\langle 0
| \tilde{B}_T | \Phi_{\mr{A}} \rangle = t_{\mr{A}}$, the $\tilde{B}_{\mr{B}}$ part eliminating
because $\langle
0_{\mr{A}} | \phi_{\mr{A}} \rangle = 0$; $\langle \Phi_{\mr{A}} | \tilde{B}_T | \Phi_{\mr{A}}
\rangle = n_{\mr{A}} ( b_{\mr{A}}^{\mr{exc}} + b_{\mr{B}} )$, the spectator contributing its ground
value; and $\langle \Phi_{\mr{A}} | \tilde{B}_T | \Phi_{\mr{B}} \rangle = 0$. Collecting terms,
\begin{equation}\label{eq:si_super_additive}
    \langle \hat{B} \rangle_{\mr{AB}}
    = \frac{1}{D} \sum_{\mr{X} = \mr{A},\mr{B}}
    \Big[ \left( |c_0|^2 + n_{\mr{Y}} \right) b_{\mr{X}}
    + n_{\mr{X}} b_{\mr{X}}^{\mr{exc}}
    + 2\, \mr{Re} \left( c_0^* t_{\mr{X}} \right) \Big]~.
    \qquad \mr{Y} \neq \mr{X}~,
\end{equation}
$\mr{X}$ contributes its ground value with weight $(|c_0|^2 + n_{\mr{Y}})/D$, the probability
that $\mr{X}$ is unexcited, its excited value with weight $n_{\mr{X}}/D$, and a coherence term.

Note that $(|c_0|^2 + n_{\mr{Y}})/D + n_{\mr{X}}/D = 1$, and Eq.~(\ref{eq:si_super_additive}) is 
of the form $\mr{Tr}(\hat{\rho}_{\mr{A}} \hat{B}_{\mr{A}}) + \mr{Tr}(\hat{\rho}_{\mr{B}}
\hat{B}_{\mr{B}})$ over the reduced states of $|\bar{\Psi}(0)\rangle$. Each subsystem sees a
normalized local distribution over its ground and excited values, with the other subsystem
traced out. This is the desired physical expression.

For example, setting $z_{\mr{B}} = 0$ gives:
\begin{equation}\label{eq:si_super_spectator}
    \langle \hat{B} \rangle_{\mr{AB}}
    = \frac{|c_0|^2 b_{\mr{A}} + n_{\mr{A}} b_{\mr{A}}^{\mr{exc}}
    + 2\,\mr{Re}\left( c_0^* t_{\mr{A}} \right)}{|c_0|^2 + n_{\mr{A}}} + b_{\mr{B}}
    = \langle \hat{B}_{\mr{A}} \rangle_{\mr{A}}^{\mr{exc}}
    + \langle \hat{B}_{\mr{B}} \rangle_{\mr{B}}^{\mr{gs}}~,
\end{equation}
the isolated $\mr{A}$ result plus the ground-state value of $\mr{B}$.

Now consider simultaneous subsystem excitation. The doubly excited component needs a generator built from
$z_{\mr{A}} z_{\mr{B}}$, which is Hermitian, so the admissible object is
\begin{equation}\label{eq:si_super_prod_def}
    \zeta = \ui\, z_{\mr{A}} z_{\mr{B}}~, \qquad \zeta^\dagger = -\zeta~,
\end{equation}
of the imaginary-symmetric type $\hat{\kappa}_{\mu + N_\omega}$ of
Eq.~(\ref{eq:si_kappa_pair}), not the real antisymmetric $\hat{\kappa}_\mu$. Take $\tilde{z}(0) =
\zeta$. Then $\langle \zeta \rangle = \ui \langle z_{\mr{A}} z_{\mr{B}} \rangle = \ui \langle
z_{\mr{A}} \rangle \langle z_{\mr{B}} \rangle = 0$, which uses the separation as well as the
convention. The weight is multiplicative,
\begin{equation}\label{eq:si_super_prod_norm}
    \langle -\zeta^2 \rangle = \langle z_{\mr{A}}^2 z_{\mr{B}}^2 \rangle
    = n_{\mr{A}} n_{\mr{B}}~, \qquad D = |c_0|^2 + n_{\mr{A}} n_{\mr{B}}~.
\end{equation}
The state has two components, $c_0|0\rangle + \ui\,|\phi_{\mr{A}}\rangle \otimes
|\phi_{\mr{B}}\rangle$. A sum of one-subsystem operators cannot change the excitation of both at
once, so the interference vanishes,
\begin{equation}\label{eq:si_super_prod_nointerf}
    \langle 0 | \tilde{B}_T |\, \phi_{\mr{A}} \phi_{\mr{B}} \rangle
    = t_{\mr{A}} \langle 0_{\mr{B}} | \phi_{\mr{B}} \rangle
    + t_{\mr{B}} \langle 0_{\mr{A}} | \phi_{\mr{A}} \rangle = 0~,
\end{equation}
and the diagonal element is $n_{\mr{A}} n_{\mr{B}} ( b_{\mr{A}}^{\mr{exc}} +
b_{\mr{B}}^{\mr{exc}} )$. Hence
\begin{equation}\label{eq:si_super_prod_result}
    \langle \hat{B} \rangle_{\mr{AB}}
    = \frac{|c_0|^2 \left( b_{\mr{A}} + b_{\mr{B}} \right)
    + n_{\mr{A}} n_{\mr{B}} \left( b_{\mr{A}}^{\mr{exc}} + b_{\mr{B}}^{\mr{exc}} \right)}
    {|c_0|^2 + n_{\mr{A}} n_{\mr{B}}}~,
\end{equation}
is a two-configuration average. One joint weight $n_{\mr{A}} n_{\mr{B}}$ now multiplies both
subsystems instead of one weight each, making $z_{\mr{B}} \to 0$ eliminates $\zeta$ and returns
$b_{\mr{A}} + b_{\mr{B}}$, so $\mr{A}$ cannot be excited alone. The reduced states are again mixed,
so this is not a sum of isolated subsystem results; only at $c_0 = 0$, where both subsystems are
excited. Eq.~(\ref{eq:si_super_prod_result}) reduce to $b_{\mr{A}}^{\mr{exc}} +
b_{\mr{B}}^{\mr{exc}}$. For $m$ subsystems the weight involves $\prod_i n_i$, exponential in $m$, against
the additive $\sum_i n_i$ of Eq.~(\ref{eq:si_super_norm}), simultaneous excitations require high
order operations.

A general initial superposition uses all four components,
\begin{equation}\label{eq:si_super_full}
    \tilde{z}(0) = \alpha_{\mr{A}} z_{\mr{A}} + \alpha_{\mr{B}} z_{\mr{B}} + \beta \zeta~,
\end{equation}
with $\alpha_{\mr{A}}$, $\alpha_{\mr{B}}$, $\beta$ real, the three generators being anti-Hermitian
and independent. The cross terms vanish, $\langle z_{\mr{A}} z_{\mr{B}} \rangle = \langle z_{\mr{A}}
\zeta \rangle = \langle z_{\mr{B}} \zeta \rangle = 0$, so $\langle \tilde{z}(0) \rangle = 0$ and
\begin{equation}\label{eq:si_super_full_norm}
    D = |c_0|^2 + \alpha_{\mr{A}}^2 n_{\mr{A}} + \alpha_{\mr{B}}^2 n_{\mr{B}}
    + \beta^2 n_{\mr{A}} n_{\mr{B}}~.
\end{equation}
This state is entangled for generic amplitudes, where two excited subsystems are
correlated. Entanglement is described here because for an additive $\hat{B}$ a normalized
state gives $\mr{Tr}(\hat{\rho}_{\mr{A}} \hat{B}_{\mr{A}}) + \mr{Tr}(\hat{\rho}_{\mr{B}}
\hat{B}_{\mr{B}})$ over its reduced states. This requires, however, that
Eq.~(\ref{eq:si_obs_start}) be an expectation value of a normalized state, where $D$ has to be the
squared norm. Eq.~(\ref{eq:si_super_full_norm}) is so, due to the requirement $\langle
\tilde{z}(0) \rangle = 0$.

Role of the reference-component convention: Each $z_{\mr{X}}$ is
free up to an anti-Hermitian constant as the equations of motion involve
only commutators. Shifting the factors first, as the convention prescribes, and forming the product
afterwards gives
\begin{equation}\label{eq:si_super_shift}
    \ui \left( z_{\mr{A}} - \langle z_{\mr{A}} \rangle \right)
    \left( z_{\mr{B}} - \langle z_{\mr{B}} \rangle \right)
    = \zeta - \ui \langle z_{\mr{B}} \rangle z_{\mr{A}}
    - \ui \langle z_{\mr{A}} \rangle z_{\mr{B}}
    + \ui \langle z_{\mr{A}} \rangle \langle z_{\mr{B}} \rangle~,
\end{equation}
whereas forming the product first and shifting it afterwards gives
\begin{equation}\label{eq:si_super_zeta_shift}
    \zeta - \langle \zeta \rangle \hat{1}
    = \zeta - \ui \langle z_{\mr{A}} z_{\mr{B}} \rangle \hat{1}~.
\end{equation}
These are different operators. They differ by the single-excitation generators $-\ui \langle
z_{\mr{B}} \rangle z_{\mr{A}} - \ui \langle z_{\mr{A}} \rangle z_{\mr{B}}$. Shifting $z_{\mr{A}}$
and $z_{\mr{B}}$ satisfies the condition $\langle \hat{\zeta}\rangle = 0$, but shifting $\zeta$
separately, using un-shifted $z_{\mr{X}}$, can be problematic.

The convention is a shift by an averaged element. A shift of a factor does not shift the product as
needed. Whichever prescription is available depends on how the manifold is
closed. If the algebra contains the doubly excited generator as a member, $\zeta$ is an eigenvector
and Eq.~(\ref{eq:si_super_zeta_shift}) applies to it directly; if $\zeta$ is only reachable as a
product of two vectors, Eq.~(\ref{eq:si_super_shift}) is the rule. For these reasons $\langle
\tilde{z}(0) \rangle = 0$ is a condition on the generator, and for a product generator it has to be
imposed on the individual operators for additivity.

The argument assumes $z_{\mr{A}}$ and $z_{\mr{B}}$ lie separately in the algebraic space. As noted
in the main text, a closed $\mathfrak{su}(M)$ need not be local, and a truncation that mixes the
subsystems will induce small nonlocality leaks.

\section{Truncated Excitation Spaces}\label{sec:truncated_excitation_spaces}

The results reported in the main text employ the full excitation space of the six-site hard-core
boson model, for which the cluster expansion terminates at excitation order 6 and the propagation is
equivalent, in principle, to FCI. To assess how the formalism behaves when the excitation space is
truncated, we repeat the driven-dynamics simulations for dynamical excitations up to order 2, 3, 4, and 6,
keeping the Hamiltonian, the initial linear combination, and the pulse parameters unchanged. In each
case the TDUCC result is compared against the exact propagation restricted to the same truncated
space, so that the comparison isolates the error of the UCC method from the error
incurred by discarding higher excitations. The excellent agreement between CI and UCC is due to the
nature of the Hamiltonian involving excitation/de-excitation operators, and 
where $\hat{b}^{\dagger}|1\rangle = 0$. In more general settings noticeable
differences between the two are expected.

\begin{figure}[htp!]
\centering
\includegraphics[scale=0.55]{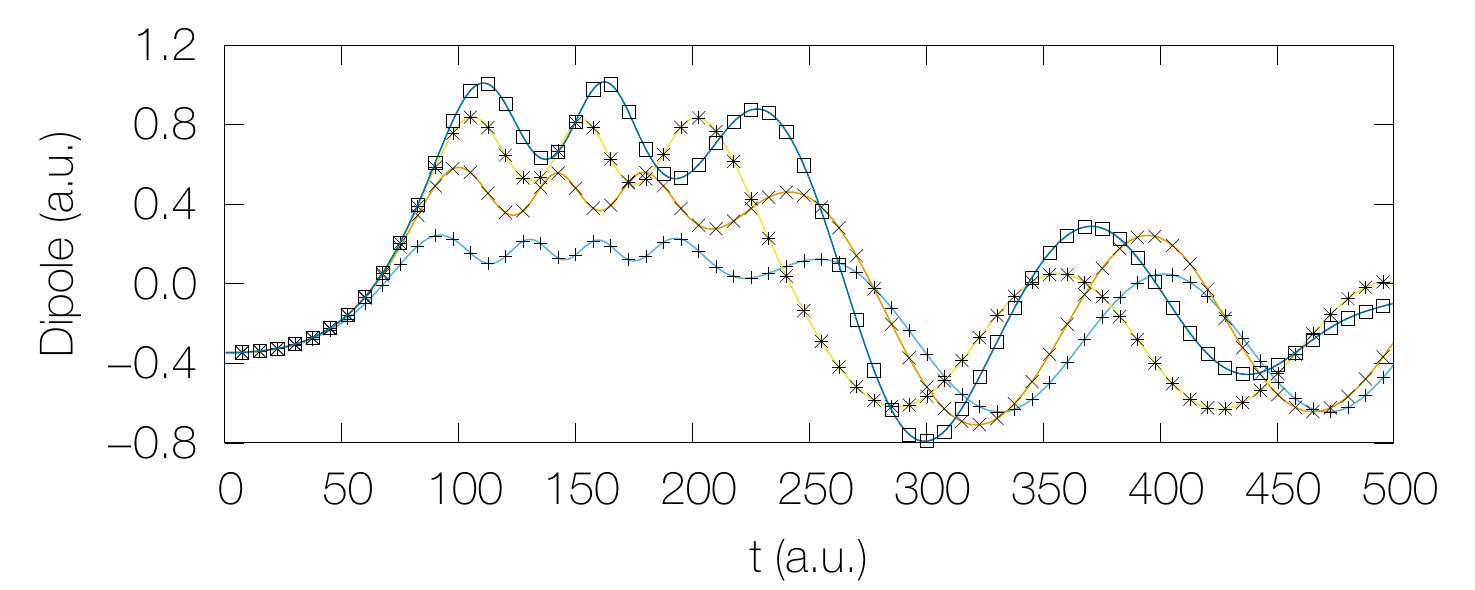}
\caption{Time-dependent dipole of the six-site hard-core boson model computed with the excitation
space truncated at order 2 (sky blue line, ``$+$'' symbols), 3 (orange line, ``$\times$''
symbols), 4 (yellow line, ``$*$'' symbols), and 6 (blue line, open-square symbols). Lines are the TDUCC
results and points/symbols are the matching CI propagation within the same truncated space; order 6 spans the
complete excitation manifold, so its reference is FCI.}
\label{fig:td_dipole_order_scan}
\end{figure}

\begin{figure}[htp!]
\centering
\includegraphics[scale=0.55]{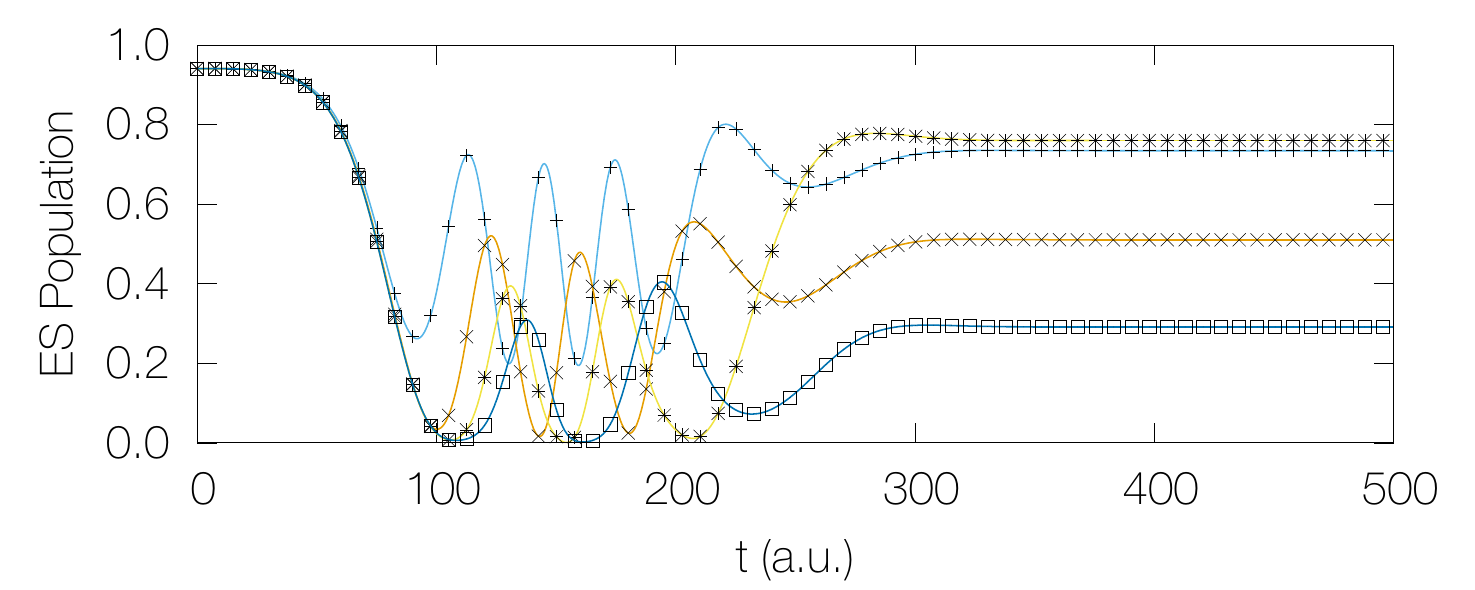}
\caption{Excited-state populations for the same excitation-order scan as
Fig.~\ref{fig:td_dipole_order_scan}: truncation at order 2 (sky blue line, ``$+$'' symbols),
3 (orange line, ``$\times$'' symbols), 4 (yellow line, ``$*$'' symbols), and 6 (blue line, open square
symbols), with TDUCC shown as lines and points/symbosl are the matching CI propagation, as in the
previous figure.}
\label{fig:ES_pops_order_scan}
\end{figure}

Figures~\ref{fig:td_dipole_order_scan} and~\ref{fig:ES_pops_order_scan} show the time-dependent
dipole and the excited-state population ($|C_1(t)|^2$), respectively. The results indicate that the truncation is
faithful to the excitation space it is given. The errors are due to the truncation itself: the
order-2 space is too small to reproduce either the dipole envelope or the post-pulse population
plateaus, while orders 3 and 4 progressively approach the untruncated (order-6, FCI) result. These
plots evidence the mid-to-strong character of the ring correlation.

\subsection{Active space and CAS-CI eigenvectors}\label{sec:active_space}

The simulations are initiated from states obtained by diagonalizing the Hamiltonian in a complete
active space built from the no-boson and the six singly-occupied configurations of the ring. Writing
each configuration as the occupation string $|n_1 n_2 n_3 n_4 n_5 n_6\rangle$ of the hard-core
bosons, the active space is spanned, in the order used hereafter, by
\begin{equation}\label{eq:si_active_space}
    \big\{\, \texttt{000000},\ \texttt{100000},\ \texttt{010000},\ \texttt{001000},\
    \texttt{000100},\ \texttt{000010},\ \texttt{000001} \,\big\}~,
\end{equation}
that is, the reference $|\Phi_0\rangle = |\texttt{000000}\rangle$ followed by the six configurations
carrying a single excitation on site $p = 1,\dots,6$.

The resulting CAS-CI eigenvectors are displayed in Table~\ref{tab:cas_ci_vectors}, one row per
eigenvector, with the columns ordered as in Eq.~(\ref{eq:si_active_space}).  Only two eigenvectors,
$K=1$ and $K=7$, carry any amplitude on the no-boson reference: they are a $2\times2$
rotation. The
remaining five eigenvectors, $K=2,\dots,6$, have strictly zero no-boson reference amplitude and span
the orthogonal complement of $|S\rangle$ inside the single-excitation space. Their coefficient
patterns are the combinations of the six-site ring. $K=2$ alternates in sign from site
to site, while $(K=3,4)$ and $(K=5,6)$ form pairs with three-site and six-site periodicity,
respectively. Because each pair is degenerate, the particular linear combination returned within a
pair is arbitrary; only the two-dimensional subspace it spans is determined.

\begin{table}[htp!]
\centering
\small
\setlength{\tabcolsep}{6pt}
\caption{CAS-CI eigenvectors of the six-site hard-core boson model, given as row vectors in the
active-space basis of Eq.~(\ref{eq:si_active_space}). Coefficients are rounded to four decimals;
entries reported by the diagonalization at $10^{-15}$ magnitude or below are reported as $0$. 
Each row is normalized to unity.}
\begin{tabular}{c rrrrrrr}
\toprule
& \multicolumn{7}{c}{Active-space configuration} \\
\cmidrule(l){2-8}
$K$ & \texttt{000000} & \texttt{100000} & \texttt{010000} & \texttt{001000} & \texttt{000100} & \texttt{000010} & \texttt{000001} \\
\midrule
1 & $-0.9345$ & $\phantom{-}0.1453$ & $\phantom{-}0.1453$ & $\phantom{-}0.1453$ & $\phantom{-}0.1453$ & $\phantom{-}0.1453$ & $\phantom{-}0.1453$ \\
2 & $\phantom{-}0$ & $\phantom{-}0.4082$ & $-0.4082$ & $\phantom{-}0.4082$ & $-0.4082$ & $\phantom{-}0.4082$ & $-0.4082$ \\
3 & $\phantom{-}0$ & $-0.5774$ & $\phantom{-}0.2887$ & $\phantom{-}0.2887$ & $-0.5774$ & $\phantom{-}0.2887$ & $\phantom{-}0.2887$ \\
4 & $\phantom{-}0$ & $\phantom{-}0$ & $\phantom{-}0.5000$ & $-0.5000$ & $\phantom{-}0$ & $\phantom{-}0.5000$ & $-0.5000$ \\
5 & $\phantom{-}0$ & $\phantom{-}0.5774$ & $\phantom{-}0.2887$ & $-0.2887$ & $-0.5774$ & $-0.2887$ & $\phantom{-}0.2887$ \\
6 & $\phantom{-}0$ & $\phantom{-}0$ & $\phantom{-}0.5000$ & $\phantom{-}0.5000$ & $\phantom{-}0$ & $-0.5000$ & $-0.5000$ \\
7 & $\phantom{-}0.3559$ & $\phantom{-}0.3815$ & $\phantom{-}0.3815$ & $\phantom{-}0.3815$ & $\phantom{-}0.3815$ & $\phantom{-}0.3815$ & $\phantom{-}0.3815$ \\
\bottomrule
\end{tabular}
\label{tab:cas_ci_vectors}
\end{table}

\subsection{Ground-state and excitation energies}\label{sec:energies}

We now compare the UCC and CI eigenvalue spectra order by order. Truncating the excitation manifold
at order $m$ involves $\sum_{k=0}^{m} \binom{6}{k}$ configurations, so the four orders appearing in
Figs.~\ref{fig:td_dipole_order_scan} and~\ref{fig:ES_pops_order_scan} support $64$, $57$, $42$ and
$22$ states, and hence $63$, $56$, $41$ and $21$ excitation energies, at orders $6$, $4$, $3$ and
$2$ respectively. Ground-state energies for all four orders are shown in
Table~\ref{tab:gs_energies}; the excitation energies $\omega_n = E_n - E_0$ are given in
Tables~\ref{tab:exc_energies_order6}, \ref{tab:exc_energies_order4}, \ref{tab:exc_energies_order3}
and \ref{tab:exc_energies_order2}. The two error scales are summarized across orders in
Table~\ref{tab:error_summary}. All energies are in atomic units; the eigenvalues are real to within
the imaginary residue of the diagonalization, which is discarded.

We consider order 6 case. Here the unitary CC method should be ``numerically exact''. This is the
reference point against which the truncated spaces are read. The agreement is indeed at the level of
double-precision arithmetic, as expected. The ground-state energies differ by $5 \times 10^{-18}$
a.u. For the excitation energies, each entry of the UCC and FCI columns of
Table~\ref{tab:exc_energies_order6} agree in the twelve printed decimals. The residuals listed in
the $\Delta_n$ columns are below $3 \times 10^{-14}$ a.u., show a mean absolute deviation of $2
\times 10^{-14}$ a.u.\ and a maximum relative deviation of $2 \times 10^{-13}$. The residuals grow
slowly with $\omega_n$, consistent with a deviation that scales with the magnitude of the root, and
are uniformly negative. We also note the degeneracy pattern of the ring. This is already visible in
the CAS-CI vectors of Table~\ref{tab:cas_ci_vectors}. Forty of the sixty-three roots group into
twenty degenerate pairs ($n = 2,3$; $4,5$; $8,9$; $10,11$; and so on), with splittings of at most $4
\times 10^{-15}$ a.u. (numerically exact degeneracies).

\begin{table}[htp!]
\centering
\small
\caption{Ground-state energy of the six-member hard-core boson model as a function of the excitation
order, computed by CI and by the unitary coupled-cluster theory, in atomic units. At
order 6 the CI result is FCI. The residual $\Delta_0 = E_0^{\mr{UCC}} - E_0^{\mr{CI}}$ measures the
error of the unitary theory within a given space; the last column is the truncation
error relative to FCI.}
\begin{tabular}{c rr r r}
\toprule
Order & $E_0^{\mr{CI}}$ & $E_0^{\mr{UCC}}$ & $\Delta_0$ & $E_0^{\mr{CI}} - E_0^{\mr{FCI}}$ \\
\midrule
6 & $-0.007702889420812862$ & $-0.007702889420812867$ & $-5.2 \times 10^{-18}$ & $0$ \\
4 & $-0.007691355150483798$ & $-0.007691355150483777$ & $+2.1 \times 10^{-17}$ & $1.153 \times 10^{-5}$ \\
3 & $-0.007665909340320160$ & $-0.007665909340320138$ & $+2.3 \times 10^{-17}$ & $3.698 \times 10^{-5}$ \\
2 & $-0.007629486442738102$ & $-0.007629486442738093$ & $+8.7 \times 10^{-18}$ & $7.340 \times 10^{-5}$ \\
\bottomrule
\end{tabular}
\label{tab:gs_energies}
\end{table}

\begin{table}[htp!]
\centering
\footnotesize
\setlength{\tabcolsep}{3pt}
\caption{Complete set of $63$ excitation energies $\omega_n = E_n - E_0$ of the six-site hard-core
boson model at excitation order 6, from FCI and from the unitary coupled-cluster theory,
in atomic units and rounded to twelve decimals. The residual
$\Delta_n = \omega_n^{\mr{UCC}} - \omega_n^{\mr{FCI}}$ is given in units of $10^{-15}$ a.u.\ (so an
entry of $-15.6$ means $-1.56 \times 10^{-14}$ a.u.). The single visible discrepancy in the printed
digits, at $n = 12$, is a rounding artifact of the twelfth decimal. Imaginary parts of the UCC
eigenvalues, of order $10^{-17}$ or smaller, are discarded.}
\begin{tabular}{r rr r @{\hspace{1.8em}} r rr r}
\toprule
$n$ & $\omega_n^{\mr{FCI}}$ & $\omega_n^{\mr{UCC}}$ & $\Delta_n$ &
$n$ & $\omega_n^{\mr{FCI}}$ & $\omega_n^{\mr{UCC}}$ & $\Delta_n$ \\
\midrule
 1 & 0.035639072276 & 0.035639072276 & $-5.0$ & 33 & 0.138957290952 & 0.138957290952 & $-18.3$ \\
 2 & 0.037803639359 & 0.037803639359 & $-6.9$ & 34 & 0.141044348669 & 0.141044348669 & $-18.3$ \\
 3 & 0.037803639359 & 0.037803639359 & $-5.1$ & 35 & 0.141547264915 & 0.141547264915 & $-19.2$ \\
 4 & 0.042102785851 & 0.042102785851 & $-8.8$ & 36 & 0.154589469637 & 0.154589469637 & $-15.8$ \\
 5 & 0.042102785851 & 0.042102785851 & $-6.0$ & 37 & 0.154617667861 & 0.154617667861 & $-15.7$ \\
 6 & 0.044236244873 & 0.044236244873 & $-8.4$ & 38 & 0.154617667861 & 0.154617667861 & $-15.2$ \\
 7 & 0.073382772583 & 0.073382772583 & $-11.7$ & 39 & 0.154726633758 & 0.154726633758 & $-16.3$ \\
 8 & 0.076694469038 & 0.076694469038 & $-13.1$ & 40 & 0.154726633758 & 0.154726633758 & $-15.7$ \\
 9 & 0.076694469038 & 0.076694469038 & $-12.0$ & 41 & 0.154839897523 & 0.154839897523 & $-15.6$ \\
10 & 0.079088088628 & 0.079088088628 & $-14.2$ & 42 & 0.187733022369 & 0.187733022369 & $-19.2$ \\
11 & 0.079088088628 & 0.079088088628 & $-11.6$ & 43 & 0.190870316913 & 0.190870316913 & $-21.2$ \\
12 & 0.079764732248 & 0.079764732247 & $-13.2$ & 44 & 0.190870316913 & 0.190870316913 & $-20.9$ \\
13 & 0.082759487089 & 0.082759487089 & $-14.5$ & 45 & 0.193208867783 & 0.193208867783 & $-22.8$ \\
14 & 0.082759487089 & 0.082759487089 & $-13.4$ & 46 & 0.193208867783 & 0.193208867783 & $-19.9$ \\
15 & 0.085271809083 & 0.085271809083 & $-14.7$ & 47 & 0.193910536658 & 0.193910536658 & $-20.8$ \\
16 & 0.097012748507 & 0.097012748507 & $-10.4$ & 48 & 0.196578112059 & 0.196578112059 & $-22.6$ \\
17 & 0.097255876154 & 0.097255876154 & $-11.7$ & 49 & 0.196578112059 & 0.196578112059 & $-22.1$ \\
18 & 0.097255876154 & 0.097255876154 & $-10.3$ & 50 & 0.198941190711 & 0.198941190711 & $-20.9$ \\
19 & 0.097722452665 & 0.097722452665 & $-11.9$ & 51 & 0.211377571086 & 0.211377571086 & $-19.9$ \\
20 & 0.097722452665 & 0.097722452665 & $-10.5$ & 52 & 0.211623556296 & 0.211623556296 & $-20.9$ \\
21 & 0.098072969035 & 0.098072969035 & $-11.9$ & 53 & 0.211623556296 & 0.211623556296 & $-18.7$ \\
22 & 0.117780046037 & 0.117780046037 & $-18.0$ & 54 & 0.212085095499 & 0.212085095499 & $-21.1$ \\
23 & 0.118383840712 & 0.118383840712 & $-19.0$ & 55 & 0.212085095499 & 0.212085095499 & $-18.3$ \\
24 & 0.132656889421 & 0.132656889421 & $-16.6$ & 56 & 0.212407504732 & 0.212407504732 & $-18.1$ \\
25 & 0.134012605772 & 0.134012605772 & $-17.2$ & 57 & 0.264365763319 & 0.264365763319 & $-23.9$ \\
26 & 0.134324721899 & 0.134324721899 & $-16.6$ & 58 & 0.266400646009 & 0.266400646009 & $-24.6$ \\
27 & 0.134324721899 & 0.134324721899 & $-16.2$ & 59 & 0.266400646009 & 0.266400646009 & $-22.8$ \\
28 & 0.134758151421 & 0.134758151421 & $-17.1$ & 60 & 0.270464290902 & 0.270464290902 & $-23.6$ \\
29 & 0.134758151421 & 0.134758151421 & $-16.9$ & 61 & 0.270464290902 & 0.270464290902 & $-23.3$ \\
30 & 0.138415638280 & 0.138415638280 & $-17.6$ & 62 & 0.272493593954 & 0.272493593954 & $-20.8$ \\
31 & 0.138415638280 & 0.138415638280 & $-17.7$ & 63 & 0.343085451979 & 0.343085451979 & $-27.0$ \\
32 & 0.138957290952 & 0.138957290952 & $-18.5$ & & & & \\
\bottomrule
\end{tabular}
\label{tab:exc_energies_order6}
\end{table}

For order 4, truncating at quadruple excitations removes seven of the sixty-four
configurations, resulting in $57$ states and $56$ excitation energies. The the UCC--CI residual is
unchanged by the truncation. The ground-state energies of Table~\ref{tab:gs_energies} differ by $2
\times 10^{-17}$ a.u., and the excitation-energy residuals of Table~\ref{tab:exc_energies_order4}
are below $2 \times 10^{-14}$ a.u., showing a mean absolute deviation of $10^{-14}$ a.u. These are the
same magnitudes found at order 6. The unitary method reproduces the CI spectrum down to the
numerical ``floor'' of the calculation. Truncating the space does not degrade it. As expected, the
truncation changes the spectrum itself. The order-4 ground state lies $1.2 \times 10^{-5}$ a.u.\
above FCI. The excitation energies deviate from their FCI counterparts by up to $1.5 \times 10^{-3}$
a.u.\ ($0.9\%$ in relative terms), giving a mean absolute deviation of $2 \times 10^{-4}$ a.u.\ over
its $56$ roots. It can be inferred that the spread between curves of different order is truncation.
We note that the comparison to FCI matches roots by ascending order, and that the two orderings are
not entirely in correspondence. The $57$-state spectrum has a degenerate pair at $n = 25,26$ where
the full spectrum has a pair at $n = 26,27$, so individual assignments near that crossing should be
read with care.

\begin{table}[htp!]
\centering
\footnotesize
\setlength{\tabcolsep}{3pt}
\caption{The $56$ excitation energies for order 4; same as before, from CI and from the unitary
coupled-cluster theory, in atomic units, rounded to twelve decimals. Every
pair of CI and UCC entries agrees in all twelve printed decimals. Imaginary parts of the UCC
eigenvalues, of order $10^{-17}$ or smaller, are discarded.}
\begin{tabular}{r rr r @{\hspace{1.8em}} r rr r}
\toprule
$n$ & $\omega_n^{\mr{CI}}$ & $\omega_n^{\mr{UCC}}$ & $\Delta_n$ &
$n$ & $\omega_n^{\mr{CI}}$ & $\omega_n^{\mr{UCC}}$ & $\Delta_n$ \\
\midrule
 1 & 0.035630978317 & 0.035630978317 & $-9.1$ & 29 & 0.134748265907 & 0.134748265907 & $-11.8$ \\
 2 & 0.037796106615 & 0.037796106615 & $-10.0$ & 30 & 0.138526425911 & 0.138526425911 & $-13.4$ \\
 3 & 0.037796106615 & 0.037796106615 & $-9.8$ & 31 & 0.138526425911 & 0.138526425911 & $-12.8$ \\
 4 & 0.042097047198 & 0.042097047198 & $-12.1$ & 32 & 0.139434826257 & 0.139434826257 & $-12.1$ \\
 5 & 0.042097047198 & 0.042097047198 & $-10.3$ & 33 & 0.139434826257 & 0.139434826257 & $-11.4$ \\
 6 & 0.044346532931 & 0.044346532931 & $-7.0$ & 34 & 0.141032814399 & 0.141032814399 & $-13.4$ \\
 7 & 0.073375695385 & 0.073375695385 & $-11.1$ & 35 & 0.142830527643 & 0.142830527643 & $-15.3$ \\
 8 & 0.076683543959 & 0.076683543959 & $-12.4$ & 36 & 0.154590197620 & 0.154590197620 & $-10.8$ \\
 9 & 0.076683543959 & 0.076683543959 & $-12.0$ & 37 & 0.154608254018 & 0.154608254018 & $-11.3$ \\
10 & 0.079087986878 & 0.079087986878 & $-13.1$ & 38 & 0.154608254018 & 0.154608254018 & $-10.5$ \\
11 & 0.079087986878 & 0.079087986878 & $-12.4$ & 39 & 0.155049313761 & 0.155049313761 & $-11.0$ \\
12 & 0.079798179605 & 0.079798179605 & $-11.5$ & 40 & 0.155049313761 & 0.155049313761 & $-11.0$ \\
13 & 0.082799743038 & 0.082799743038 & $-12.9$ & 41 & 0.155949151467 & 0.155949151467 & $-13.8$ \\
14 & 0.082799743038 & 0.082799743038 & $-13.0$ & 42 & 0.187756341351 & 0.187756341351 & $-12.1$ \\
15 & 0.085518509004 & 0.085518509004 & $-13.1$ & 43 & 0.190872073956 & 0.190872073956 & $-12.4$ \\
16 & 0.097001214237 & 0.097001214237 & $-11.1$ & 44 & 0.190872073956 & 0.190872073956 & $-12.5$ \\
17 & 0.097247413055 & 0.097247413055 & $-11.3$ & 45 & 0.193368838378 & 0.193368838378 & $-12.4$ \\
18 & 0.097247413055 & 0.097247413055 & $-10.2$ & 46 & 0.193368838378 & 0.193368838378 & $-10.6$ \\
19 & 0.097755416307 & 0.097755416307 & $-10.3$ & 47 & 0.194868197755 & 0.194868197755 & $-5.0$ \\
20 & 0.097755416307 & 0.097755416307 & $-10.2$ & 48 & 0.197328164116 & 0.197328164116 & $-8.6$ \\
21 & 0.098366465503 & 0.098366465503 & $-9.5$ & 49 & 0.197328164116 & 0.197328164116 & $-7.8$ \\
22 & 0.117946949558 & 0.117946949558 & $-14.5$ & 50 & 0.200411258396 & 0.200411258396 & $-12.6$ \\
23 & 0.118482871746 & 0.118482871746 & $-14.0$ & 51 & 0.211366036816 & 0.211366036816 & $-11.0$ \\
24 & 0.132645355150 & 0.132645355150 & $-11.5$ & 52 & 0.211638813651 & 0.211638813651 & $-11.8$ \\
25 & 0.134314048756 & 0.134314048756 & $-12.3$ & 53 & 0.211638813651 & 0.211638813651 & $-9.3$ \\
26 & 0.134314048756 & 0.134314048756 & $-11.5$ & 54 & 0.212347552103 & 0.212347552103 & $-7.1$ \\
27 & 0.134336779305 & 0.134336779305 & $-10.5$ & 55 & 0.212347552103 & 0.212347552103 & $-5.7$ \\
28 & 0.134748265907 & 0.134748265907 & $-12.6$ & 56 & 0.213769170474 & 0.213769170474 & $-6.8$ \\
\bottomrule
\end{tabular}
\label{tab:exc_energies_order4}
\end{table}

For order 3, truncating at triple excitations leaves $42$ states and $41$ excitation
energies. The trends established at order 4 are confirmed without change. The UCC and CI
ground-state energies differ by $2 \times 10^{-17}$ a.u., and the excitation-energy residuals of
Table~\ref{tab:exc_energies_order3} stay below $2 \times 10^{-14}$ a.u., with a mean absolute
deviation of $10^{-14}$ a.u., indistinguishable from the orders 4 and 6 figures. Across the three
orders the method error is flat, at roughly $10^{-14}$ a.u., and shows no dependence on
the size of the space.

The truncation error, by contrast, grows monotonically as excitations are removed. The order-3
ground state lies $3.7 \times 10^{-5}$ a.u.\ above FCI, three times the order-4 value, and the
excitation energies depart from their FCI counterparts by up to $1.6 \times 10^{-3}$ a.u.\
($1.3\%$), with a mean absolute deviation of $3 \times 10^{-4}$ a.u.\ over its $41$ roots. A
comparison against order 4 is shown in Table~\ref{tab:error_summary}, since the two spaces support
different numbers of roots and the deviations grow with $\omega_n$. The ordering caveat noted for
order 4 applies here as well.  The order-3 spectrum has a degenerate pair at $n = 25,26$ where the
full spectrum has one at $n = 26,27$.

\begin{table}[htp!]
\centering
\footnotesize
\setlength{\tabcolsep}{3pt}
\caption{The $41$ excitation energies for order 3 truncation.}
\begin{tabular}{r rr r @{\hspace{1.8em}} r rr r}
\toprule
$n$ & $\omega_n^{\mr{CI}}$ & $\omega_n^{\mr{UCC}}$ & $\Delta_n$ &
$n$ & $\omega_n^{\mr{CI}}$ & $\omega_n^{\mr{UCC}}$ & $\Delta_n$ \\
\midrule
 1 & 0.035620724740 & 0.035620724740 & $-10.8$ & 22 & 0.118281264884 & 0.118281264884 & $-15.2$ \\
 2 & 0.037789921752 & 0.037789921752 & $-11.6$ & 23 & 0.118941842989 & 0.118941842989 & $-12.8$ \\
 3 & 0.037789921752 & 0.037789921752 & $-11.6$ & 24 & 0.132619909340 & 0.132619909340 & $-14.9$ \\
 4 & 0.042103425218 & 0.042103425218 & $-13.1$ & 25 & 0.134425608834 & 0.134425608834 & $-11.1$ \\
 5 & 0.042103425218 & 0.042103425218 & $-13.0$ & 26 & 0.134425608834 & 0.134425608834 & $-11.1$ \\
 6 & 0.044505628368 & 0.044505628368 & $-18.0$ & 27 & 0.134527095886 & 0.134527095886 & $-7.8$ \\
 7 & 0.073400227952 & 0.073400227952 & $-13.3$ & 28 & 0.134880339164 & 0.134880339164 & $-11.4$ \\
 8 & 0.076732033757 & 0.076732033757 & $-14.6$ & 29 & 0.134880339164 & 0.134880339164 & $-11.4$ \\
 9 & 0.076732033757 & 0.076732033757 & $-14.6$ & 30 & 0.138724196926 & 0.138724196926 & $-14.5$ \\
10 & 0.079200708352 & 0.079200708352 & $-14.8$ & 31 & 0.138724196926 & 0.138724196926 & $-14.5$ \\
11 & 0.079200708352 & 0.079200708352 & $-14.8$ & 32 & 0.139790130672 & 0.139790130672 & $-6.4$ \\
12 & 0.080176223708 & 0.080176223708 & $-13.4$ & 33 & 0.139790130672 & 0.139790130672 & $-6.5$ \\
13 & 0.083216450819 & 0.083216450819 & $-15.2$ & 34 & 0.141410910021 & 0.141410910021 & $-12.0$ \\
14 & 0.083216450819 & 0.083216450819 & $-15.1$ & 35 & 0.143132023529 & 0.143132023529 & $-9.6$ \\
15 & 0.086122860373 & 0.086122860373 & $-15.6$ & 36 & 0.154605921911 & 0.154605921911 & $-13.7$ \\
16 & 0.097004276944 & 0.097004276944 & $-12.6$ & 37 & 0.154667730712 & 0.154667730712 & $-11.0$ \\
17 & 0.097322335876 & 0.097322335876 & $-12.7$ & 38 & 0.154667730712 & 0.154667730712 & $-10.9$ \\
18 & 0.097322335876 & 0.097322335876 & $-12.8$ & 39 & 0.155216518009 & 0.155216518009 & $-7.0$ \\
19 & 0.098154908432 & 0.098154908432 & $-12.2$ & 40 & 0.155216518009 & 0.155216518009 & $-6.3$ \\
20 & 0.098154908432 & 0.098154908432 & $-12.2$ & 41 & 0.156182776365 & 0.156182776365 & $-6.8$ \\
21 & 0.099386274545 & 0.099386274545 & $-11.5$ & & & & \\
\bottomrule
\end{tabular}
\label{tab:exc_energies_order3}
\end{table}

Let us consider order 2. The smallest space considered keeps only the reference, the six single and
the fifteen double excitations, giving $22$ states and $21$ excitation energies. The UCC
error again does not change. The ground-state energies differ by $9 \times 10^{-18}$ a.u., and the
residuals of Table~\ref{tab:exc_energies_order2} are below
$1.5 \times 10^{-14}$ a.u.; mean absolute deviation of $9 \times 10^{-15}$ a.u. The
truncation error, on the other hand, is now at its largest. The ground state is
$7.3 \times 10^{-5}$ a.u.\ above FCI, and the excitation energies are displaced by up to
$1.5 \times 10^{-3}$ a.u., which at these low-lying frequencies amounts to $2.8\%$. In contrast to the
larger spaces, each order-2 excitation energy is above its FCI counterpart. The
degeneracy pattern of the lowest $21$ roots matches the full spectrum, so no ordering
caveat is needed here.

Table~\ref{tab:error_summary} shows the two error scales across all four orders. The
truncation error evaluated over the lowest $21$ roots so that the four spaces are compared on a
common set of states. It makes the central point of this section quantitative. The
UCC error is flat at $\sim 10^{-14}$ a.u.\ over spaces ranging from $22$ to $64$
states. The truncation error, in contrast, rises by an order of magnitude from order 4 to
order 2, from $0.3\%$ to $2.8\%$ in the worst case. The two scales are separated by ten orders of
magnitude, which is why the curves of Figs.~\ref{fig:td_dipole_order_scan}
and~\ref{fig:ES_pops_order_scan} may be read as a measure of excitation-space truncation.

The excitation basis results from decomposition of their metric $\tilde{S}_{kl} = \langle 0 |
\hat{\tau}_k^\dagger \hat{\tau}_l | 0 \rangle$, keeping those singular vectors whose singular values
$s$ exceed a threshold $\eta = 10^{-6}$ and rescaling them by $s^{-1/2}$. In all cases reported in
this section, covering all four excitation orders, the singular values are between $0.3$ and $1.9$.
They are not tabulated, as only this range is relevant here.  We note now that: i) The smallest
singular value is five orders of magnitude above $\eta$. No meaningful direction is discarded, and
the excitation basis is of full rank within each truncation. From the $\sum_{k=0}^{m} \binom{6}{k}$
states none is removed for linear dependence. ii) Their metric is well conditioned, with a ratio of
largest to smallest singular value of at most $7$. The $s^{-1/2}$ rescaling amplifies no component
by more than $s_{\min}^{-1/2} \approx 2$. Neither figure degrades as the space is truncated,
consistent with the uniform $10^{-14}$ a.u.\ residuals found at every order.

The dominant contribution of error is the convergence of the ground-state amplitudes. The iteration
is stopped once the amplitude step falls below $10^{-13}$, the residual error of the converged
vector can be read directly from Table~\ref{tab:t_amplitudes}. The amplitudes forbidden by the
symmetry of the reference, which should vanish identically, instead are at up to $4 \times
10^{-13}$. An amplitude error $\delta t$ enters $E_0$ at around second order. This is why the
ground-state columns of Table~\ref{tab:gs_energies} agree to $10^{-17}$--$10^{-18}$ a.u.  The
excitation energies behave differently. The Jacobian depends on the amplitudes through $\hat{H}_T =
e^{-\hat{T}} \hat{H}_0 e^{\hat{T}}$, whose first-order variation is $[\hat{H}_T, \delta \hat{T}]$,
so the induced shift is linear in the amplitude error, $\delta \Omega \sim \| \hat{H}_0 \| \, \delta
t$. With the spectrum of $\hat{H}_0$ spanning $\sim 10^{-1}$ a.u., this places the floor in the low
$10^{-14}$ a.u.\ range, as Table~\ref{tab:error_summary} reports. 

It follows that the intrinsic error of the truncated UCC method is smaller than the values
tabulated here, so the ten-order-of-magnitude separation from the truncation error can be
understated. Resolving the true figure would require tightening the amplitude convergence threshold,
which would not change conclusions drawn in this section. We note that the lowest excitation
is non-degenerate at all four orders ($\omega_1$ and $\omega_2$ are separated by $\approx 2 \times
10^{-3}$ a.u., the first degenerate pair being $n = 2,3$). The eigenvector of
Sec.~\ref{sec:eom_vector} and the initial state built from it in the time-dependent simulations are
uniquely determined.

\begin{table}[htp!]
\centering
\footnotesize
\setlength{\tabcolsep}{3pt}
\caption{$21$ excitation energies for order 2 excitation truncation.}
\begin{tabular}{r rr r @{\hspace{1.8em}} r rr r}
\toprule
$n$ & $\omega_n^{\mr{CI}}$ & $\omega_n^{\mr{UCC}}$ & $\Delta_n$ &
$n$ & $\omega_n^{\mr{CI}}$ & $\omega_n^{\mr{UCC}}$ & $\Delta_n$ \\
\midrule
 1 & 0.035782101919 & 0.035782101919 & $-9.0$ & 12 & 0.080457387974 & 0.080457387974 & $-6.6$ \\
 2 & 0.037965298549 & 0.037965298549 & $-9.7$ & 13 & 0.083298300865 & 0.083298300865 & $-7.2$ \\
 3 & 0.037965298549 & 0.037965298549 & $-9.7$ & 14 & 0.083298300865 & 0.083298300865 & $-7.0$ \\
 4 & 0.042329551518 & 0.042329551518 & $-11.2$ & 15 & 0.086295321439 & 0.086295321439 & $-10.3$ \\
 5 & 0.042329551518 & 0.042329551518 & $-11.1$ & 16 & 0.097013638081 & 0.097013638081 & $-9.9$ \\
 6 & 0.045495720897 & 0.045495720897 & $-13.8$ & 17 & 0.097401317992 & 0.097401317992 & $-8.6$ \\
 7 & 0.073451728134 & 0.073451728134 & $-9.9$ & 18 & 0.097401317992 & 0.097401317992 & $-8.7$ \\
 8 & 0.076797663930 & 0.076797663930 & $-9.8$ & 19 & 0.098288060080 & 0.098288060080 & $-5.0$ \\
 9 & 0.076797663930 & 0.076797663930 & $-9.8$ & 20 & 0.098288060080 & 0.098288060080 & $-4.9$ \\
10 & 0.079295610403 & 0.079295610403 & $-9.7$ & 21 & 0.099594878190 & 0.099594878190 & $-4.6$ \\
11 & 0.079295610403 & 0.079295610403 & $-9.7$ & & & & \\
\bottomrule
\end{tabular}
\label{tab:exc_energies_order2}
\end{table}

\begin{table}[htp!]
\centering
\small
\setlength{\tabcolsep}{5pt}
\caption{The two error scales of the excitation-order scan, in atomic units. ``States'' is the
dimension $\sum_{k=0}^{m}\binom{6}{k}$ of the truncated space and $N_\omega$ the number of
excitation energies it supports. The method error
$\Delta_n = \omega_n^{\mr{UCC}} - \omega_n^{\mr{CI}}$ is taken over all $N_\omega$ roots of the
given order. The truncation error $\delta_n = \omega_n^{\mr{CI}} - \omega_n^{\mr{FCI}}$ is
evaluated over the lowest $21$ roots only, common to all four spaces, so that the orders are
compared on the same set of states; the relative column is $\max_n |\delta_n / \omega_n^{\mr{FCI}}|$
over that set. Roots are matched by ascending order.}
\begin{tabular}{c cc rr rr r}
\toprule
& & & \multicolumn{2}{c}{Method error} & \multicolumn{3}{c}{Truncation error} \\
\cmidrule(lr){4-5} \cmidrule(l){6-8}
Order & States & $N_\omega$ & mean $|\Delta_n|$ & max $|\Delta_n|$ &
mean $|\delta_n|$ & max $|\delta_n|$ & max rel. \\
\midrule
6 & 64 & 63 & $1.7 \times 10^{-14}$ & $2.7 \times 10^{-14}$ & $0$ & $0$ & $0$ \\
4 & 57 & 56 & $1.1 \times 10^{-14}$ & $1.5 \times 10^{-14}$ & $4.4 \times 10^{-5}$ & $2.9 \times 10^{-4}$ & $0.30\%$ \\
3 & 42 & 41 & $1.2 \times 10^{-14}$ & $1.8 \times 10^{-14}$ & $2.4 \times 10^{-4}$ & $1.3 \times 10^{-3}$ & $1.34\%$ \\
2 & 22 & 21 & $8.9 \times 10^{-15}$ & $1.4 \times 10^{-14}$ & $4.1 \times 10^{-4}$ & $1.5 \times 10^{-3}$ & $2.85\%$ \\
\bottomrule
\end{tabular}
\label{tab:error_summary}
\end{table}

\subsection{Converged cluster amplitudes}\label{sec:t_amplitudes}

It is instructive to look at the converged ground-state cluster operator itself. The generator basis
is built in pairs from the orthonormalized excitation operators $\hat{\tau}_\mu$ of
Sec.~\ref{sec:energies}: for each $\mu = 1,\dots,N_\omega$ one forms a real antisymmetric generator
and an imaginary symmetric one,
\begin{equation}\label{eq:si_kappa_pair}
    \hat{\kappa}_\mu = \hat{\tau}_\mu - \hat{\tau}_\mu^{\dagger}~, \qquad
    \hat{\kappa}_{\mu + N_\omega} = \ui \left( \hat{\tau}_\mu + \hat{\tau}_\mu^{\dagger} \right)~,
\end{equation}
so that $\hat{T} = \sum_{\mu=1}^{2N_\omega} t_\mu \hat{\kappa}_\mu$. 
The amplitude vector therefore has $2N_\omega$ entries: $126$,
$112$, $82$ and $42$ at orders $6$, $4$, $3$ and $2$.

For ground state, the entire second block ($\mu + N_\omega$) vanishes identically, as exact zeros at every order. 
The static ground-state cluster operator uses only the
real antisymmetric basis operators, $e^{\hat{T}}$ is then a real orthogonal rotation; no
imaginary-symmetric component is needed. Within the non-vanishing block only about half the
amplitudes are nonzero, $33$ of $63$ at order 6, $31$ of $56$ at order 4, $21$ of $41$ at order 3
and $11$ of $21$ at order 2. The remainder are numerically vanishing,
bounded by $5 \times 10^{-13}$, regardless of order; these are amplitudes forbidden by the symmetry
of the reference. The consistency of the fraction across four differently sized spaces suggests a
symmetry selection rule. However, we did not the basis by irreducible
representations in this work.

Table~\ref{tab:t_amplitudes} lists the non-negligible amplitudes. The norm of the cluster vector is
almost independent of the truncation, $\|\mb{t}\| \approx 0.109$ at all four orders, a spread of
about one percent across spaces differing by a factor of three in size. How the fixed total is
distributed changes, however: the largest single amplitude grows monotonically as the space
contracts, from $0.069$ at order 6 to $0.108$ at order 2, where a single component carries almost
the entire norm. The cluster operator does not shrink when excitations are removed, it
concentrates instead. This is consistent with the ground-state energies of
Table~\ref{tab:gs_energies}.

\begin{table}[htp!]
\centering
\footnotesize
\setlength{\tabcolsep}{4pt}
\caption{Converged ground-state cluster amplitudes $t_\mu$ of
$\hat{T} = \sum_\mu t_\mu \hat{\kappa}_\mu$, for each excitation order. Only the amplitudes of the real
antisymmetric generators $\hat{\kappa}_\mu = \hat{\tau}_\mu - \hat{\tau}_\mu^{\dagger}$ are shown,
and from those only the terms exceeding $10^{-10}$ in magnitude. The amplitudes omitted from this block are
numerically zero (below $4.2 \times 10^{-13}$ at any order), and the amplitudes of the
imaginary-symmetric generators $\ui(\hat{\tau}_\mu + \hat{\tau}_\mu^{\dagger})$ are zero and
are not listed. The index $\mu$ labels each order's own SVD- and QR-adapted generator basis and is
not comparable between columns: the four spaces have different dimensions and their bases are
constructed independently.}
\begin{tabular}{r r @{\hspace{1.4em}} r r @{\hspace{1.4em}} r r @{\hspace{1.4em}} r r}
\toprule
\multicolumn{2}{c}{Order 6} & \multicolumn{2}{c}{Order 4} &
\multicolumn{2}{c}{Order 3} & \multicolumn{2}{c}{Order 2} \\
\cmidrule(r){1-2} \cmidrule(lr){3-4} \cmidrule(lr){5-6} \cmidrule(l){7-8}
$\mu$ & $t_\mu$ & $\mu$ & $t_\mu$ & $\mu$ & $t_\mu$ & $m$ & $t_\mu$ \\
\midrule
 1 &  0.06879744 &  6 &  0.06420044 &  6 & $-0.08886644$ &  6 & $-0.10816584$ \\
 7 &  0.02983811 &  7 & $-0.00007443$ &  7 &  0.00026546 &  7 &  0.00033215 \\
13 & $-0.00108730$ &  8 &  0.00179527 &  8 & $-0.00056626$ &  8 & $-0.00039818$ \\
14 & $-0.00091403$ &  9 &  0.00213987 &  9 &  0.00321174 &  9 & $-0.00114467$ \\
15 &  0.00070310 & 10 & $-0.00105710$ & 10 & $-0.00212255$ & 10 &  0.00361245 \\
16 & $-0.00108404$ & 11 & $-0.00229692$ & 11 &  0.00211404 & 11 & $-0.00075070$ \\
17 &  0.00005609 & 12 & $-0.00196333$ & 12 & $-0.00270004$ & 12 &  0.00189317 \\
18 &  0.00181999 & 13 & $-0.00248476$ & 13 &  0.00182995 & 13 & $-0.00053871$ \\
19 &  0.00123370 & 14 &  0.00021818 & 14 &  0.00308483 & 14 & $-0.00682671$ \\
20 &  0.00081749 & 15 & $-0.00012737$ & 15 & $-0.00146654$ & 15 &  0.00121313 \\
21 & $-0.00389021$ & 21 & $-0.07996618$ & 26 & $-0.06020010$ & 21 &  0.01240209 \\
22 & $-0.06901746$ & 22 & $-0.00075836$ & 27 &  0.00054988 & & \\
28 &  0.00225817 & 23 & $-0.00484906$ & 28 &  0.00014475 & & \\
29 & $-0.00235665$ & 24 & $-0.00033903$ & 29 & $-0.00071229$ & & \\
30 &  0.00068059 & 25 & $-0.00022298$ & 30 &  0.00199964 & & \\
31 &  0.00003163 & 26 &  0.00116132 & 31 &  0.00197928 & & \\
32 & $-0.00007583$ & 27 &  0.00268338 & 32 &  0.00102276 & & \\
33 &  0.00131390 & 28 & $-0.00060300$ & 33 &  0.00201415 & & \\
34 & $-0.00030189$ & 29 & $-0.00131597$ & 34 &  0.00348872 & & \\
35 & $-0.00190139$ & 30 & $-0.00118040$ & 35 &  0.00083230 & & \\
36 & $-0.00443156$ & 41 & $-0.00833795$ & 41 &  0.01327970 & & \\
42 &  0.01128379 & 42 &  0.00071382 & & & & \\
48 & $-0.00013957$ & 43 & $-0.00060894$ & & & & \\
49 &  0.00110990 & 44 &  0.00101409 & & & & \\
50 &  0.00142199 & 45 &  0.00017157 & & & & \\
51 & $-0.00249389$ & 46 & $-0.00087724$ & & & & \\
52 & $-0.00052648$ & 47 & $-0.00211920$ & & & & \\
53 & $-0.00009684$ & 48 &  0.00018061 & & & & \\
54 &  0.00008557 & 49 &  0.00036935 & & & & \\
55 & $-0.00150723$ & 50 &  0.00186543 & & & & \\
56 & $-0.00006557$ & 56 &  0.03603878 & & & & \\
57 &  0.02150804 & & & & & & \\
63 &  0.03157080 & & & & & & \\
\bottomrule
\end{tabular}
\label{tab:t_amplitudes}
\end{table}

\subsection{First EOM-UCC excitation vector}\label{sec:eom_vector}

The dynamics of Figs.~\ref{fig:td_dipole_order_scan} and~\ref{fig:ES_pops_order_scan} are initiated
from a linear combination of the ground state and the lowest excited state, the latter generated by
the first EOM-UCC eigenvector. Since that vector fixes the initial condition, we record it here.
The frequency it belongs to is $\approx 0.036$ a.u.\ at all four orders, given to full precision by
the $n = 1$ UCC entries of Tables~\ref{tab:exc_energies_order6}--\ref{tab:exc_energies_order2}.

The eigenvector is written $\tilde{X}^I = \sum_\nu X_\nu^I \hat{\Xi}_{0,\nu}$ with
$\hat{\Xi}_{0,\nu} = \hat{\pi}_T(\hat{\kappa}_\nu)$, so its $2N_\omega$ components are labeled by the
paired generator basis of Eq.~(\ref{eq:si_kappa_pair}).
The eigenvector effects the dynamics through the initial generator $\tilde{z}(0) = \sum_{I>0}
(c_I \tilde{X}^I - c_I^* \tilde{X}^{I\dagger})$ of the main text. Since the $\hat{\Xi}_{0,\nu}$
are anti-Hermitian, $\tilde{X}^{I\dagger} = -\sum_\nu (X_\nu^I)^* \hat{\Xi}_{0,\nu}$, the
combination antisymmetrizes to
\begin{equation}\label{eq:si_z0_real}
    \tilde{z}(0) = \sum_{I>0} \sum_\nu 2\,\mr{Re}\left( c_I X_\nu^I \right) \hat{\Xi}_{0,\nu}~.
\end{equation}
Only $\mr{Re}(c_I X_\nu^I)$ remains. Additionally, an eigenvector of the generalized problem is
fixed up to a global complex factor, which can be absorbed into $c_I$. The state
depends on the product $c_I \tilde{X}^I$, not on the two separately. 

We fix the phase by requiring the largest component on the real antisymmetric block to be real,
the residual sign by requiring positive overlap with the corresponding CI excited state, which
makes the anchoring entry ($\mu = 26$, $37$, $19$ and $1$ at orders $6$, $4$, $3$ and $2$) negative
at orders 3 and 2. Under this convention the first block, real antisymmetric generators
$\hat{\kappa}_\mu$, is purely real and the second block, imaginary-symmetric
$\hat{\kappa}_{\mu+N_\omega}$, purely imaginary, both to $1.4 \times 10^{-15}$ or better 
in general. With real $c_I$ the second block drops out of Eq.~(\ref{eq:si_z0_real}).
The first column of each panel of Table~\ref{tab:eom_vector} is the vector that builds the
initial condition. 

As with the cluster amplitudes of Sec.~\ref{sec:t_amplitudes}, roughly half of the components
vanish, $30$ of $63$, $25$ of $56$, $20$ of $41$ and $10$ of $21$ surviving, the same index set is
nonzero in both blocks. The block norms are insensitive to the truncation, $\approx 0.497$ for
the first and $\approx 0.507$ for the second at all four orders. The same behavior is already seen
in Table~\ref{tab:t_amplitudes}.

\begin{table}[htp!]
\centering
\let\tabular\oldtabular
\fontsize{7}{8.5}\selectfont
\setlength{\tabcolsep}{1pt}
\caption{First EOM-UCC excitation vector at each excitation order, in the generator basis of
Eq.~(\ref{eq:si_kappa_pair}). For each order, $\mr{Re}\,X_\mu$ is the coefficient of the real
antisymmetric generator $\hat{\kappa}_\mu = \hat{\tau}_\mu - \hat{\tau}_\mu^{\dagger}$ and
$\mr{Im}\,X_{\mu'}$, with $\mu' = \mu + N_\omega$, that of the paired imaginary-symmetric generator
$\ui(\hat{\tau}_\mu + \hat{\tau}_\mu^{\dagger})$. Only $2\,\mr{Re}(c_I X_\nu)$ enters the initial
generator, Eq.~(\ref{eq:si_z0_real}); with the phase convention used here and real $c_I$ that is the
first block alone, and the second is listed only to show that the convention is met. The first block
is real and the second purely imaginary to $1.4 \times 10^{-15}$ or better, so the two omitted parts
of each entry are ``numerical'' zeros. Only components exceeding $10^{-10}$ are shown; the same index set
is nonzero in both blocks. The phase is fixed so that the largest first-block component is real, and
the residual sign so that the state overlaps the corresponding CI state positively. The index $\mu$
labels each order's own generator basis and is not comparable between columns.}
\begin{tabular}{r rr @{\hspace{0.6em}} r rr @{\hspace{0.6em}} r rr @{\hspace{0.6em}} r rr}
\toprule
\multicolumn{3}{c}{Order 6} & \multicolumn{3}{c}{Order 4} &
\multicolumn{3}{c}{Order 3} & \multicolumn{3}{c}{Order 2} \\
\cmidrule(r){1-3} \cmidrule(lr){4-6} \cmidrule(lr){7-9} \cmidrule(l){10-12}
$\mu$ & $\mr{Re}\,X_\mu$ & $\mr{Im}\,X_{\mu'}$ &
$\mu$ & $\mr{Re}\,X_\mu$ & $\mr{Im}\,X_{\mu'}$ &
$\mu$ & $\mr{Re}\,X_\mu$ & $\mr{Im}\,X_{\mu'}$ &
$\mu$ & $\mr{Re}\,X_\mu$ & $\mr{Im}\,X_{\mu'}$ \\
\midrule
 2 &  0.07277689 & $-0.07417270$ &  1 & $-0.00984646$ &  0.01007296 &  1 &  0.05793987 & $-0.05870815$ &  1 & $-0.29387447$ &  0.30200544 \\
 3 & $-0.07897602$ &  0.08049072 &  2 &  0.07815983 & $-0.07995774$ &  2 & $-0.15877790$ &  0.16088329 &  2 & $-0.02773292$ &  0.02850024 \\
 4 & $-0.04624741$ &  0.04713440 &  3 & $-0.13034091$ &  0.13333915 &  3 &  0.14643150 & $-0.14837318$ &  3 & $-0.16318744$ &  0.16770254 \\
 5 &  0.08492923 & $-0.08655810$ &  4 & $-0.05125001$ &  0.05242892 &  4 & $-0.03224959$ &  0.03267722 &  4 &  0.10904473 & $-0.11206179$ \\
 6 & $-0.06994981$ &  0.07129139 &  5 & $-0.08364807$ &  0.08557223 &  5 &  0.11687713 & $-0.11842691$ &  5 & $-0.14488246$ &  0.14889109 \\
 8 & $-0.03136786$ &  0.03200192 & 16 & $-0.11477365$ &  0.11679319 & 16 & $-0.20867423$ &  0.21414017 & 16 & $-0.05360857$ &  0.05423198 \\
 9 &  0.15217221 & $-0.15524816$ & 17 &  0.18205084 & $-0.18525418$ & 17 & $-0.00988757$ &  0.01014656 & 17 & $-0.17171007$ &  0.17370685 \\
10 & $-0.10210256$ &  0.10416642 & 18 &  0.17804043 & $-0.18117320$ & 18 &  0.09804108 & $-0.10060913$ & 18 & $-0.11839243$ &  0.11976919 \\
11 &  0.21249670 & $-0.21679203$ & 19 &  0.19147812 & $-0.19484734$ & 19 & $-0.30719355$ &  0.31524007 & 19 & $-0.11449537$ &  0.11582682 \\
12 & $-0.04268645$ &  0.04354930 & 20 &  0.00420027 & $-0.00427418$ & 20 & $-0.02400756$ &  0.02463640 & 20 &  0.20257042 & $-0.20492608$ \\
23 &  0.06137529 & $-0.06268538$ & 31 &  0.00053351 & $-0.00056999$ & 21 & $-0.00011411$ &  0.00011442 & & & \\
24 &  0.07147988 & $-0.07300567$ & 32 & $-0.00022778$ &  0.00022659 & 22 &  0.00017395 & $-0.00017640$ & & & \\
25 & $-0.04455307$ &  0.04550408 & 33 &  0.00008556 & $-0.00010182$ & 23 &  0.00021068 & $-0.00021764$ & & & \\
26 &  0.25381339 & $-0.25923121$ & 34 & $-0.00037026$ &  0.00038923 & 24 &  0.00070650 & $-0.00072573$ & & & \\
27 & $-0.12731785$ &  0.13003553 & 35 &  0.00021728 & $-0.00022507$ & 25 &  0.00020017 & $-0.00020291$ & & & \\
37 &  0.00018969 & $-0.00019185$ & 36 &  0.03408838 & $-0.03496802$ & 36 & $-0.13847630$ &  0.13972729 & & & \\
38 &  0.00060482 & $-0.00060896$ & 37 &  0.22889866 & $-0.23480531$ & 37 &  0.07282453 & $-0.07348243$ & & & \\
39 & $-0.00013151$ &  0.00013416 & 38 &  0.11544398 & $-0.11842298$ & 38 &  0.06601567 & $-0.06661205$ & & & \\
40 & $-0.00037726$ &  0.00037597 & 39 &  0.11130147 & $-0.11417357$ & 39 &  0.04320074 & $-0.04359101$ & & & \\
41 &  0.00000548 & $-0.00000870$ & 40 &  0.08655735 & $-0.08879093$ & 40 &  0.05713357 & $-0.05764971$ & & & \\
43 & $-0.00344607$ &  0.00352173 & 51 & $-0.00969200$ &  0.00976482 & & & & & & \\
44 & $-0.12187830$ &  0.12455433 & 52 & $-0.08930124$ &  0.08997212 & & & & & & \\
45 & $-0.02667470$ &  0.02726038 & 53 & $-0.02355840$ &  0.02373538 & & & & & & \\
46 &  0.10551372 & $-0.10783044$ & 54 & $-0.04087515$ &  0.04118223 & & & & & & \\
47 &  0.12263501 & $-0.12532766$ & 55 &  0.04654853 & $-0.04689823$ & & & & & & \\
58 &  0.01461960 & $-0.01487529$ & & & & & & & & & \\
59 &  0.01635633 & $-0.01664239$ & & & & & & & & & \\
60 &  0.03349750 & $-0.03408335$ & & & & & & & & & \\
61 & $-0.03944699$ &  0.04013688 & & & & & & & & & \\
62 &  0.05438044 & $-0.05533152$ & & & & & & & & & \\
\bottomrule
\end{tabular}
\label{tab:eom_vector}
\end{table}

\end{document}